\documentclass[manuscript]{acmart}
\usepackage{multirow}
\usepackage{amsmath}

\setcopyright{acmlicensed}
\copyrightyear{2024}
\acmYear{2024}
\acmDOI{XXXXXXX.XXXXXXX}
\acmISBN{978-1-4503-XXXX-X/18/06}

\begin{document}



\title{IoT Firmware Version Identification Using Transfer Learning with Twin Neural Networks}

\author{Ashley Andrews}
\orcid{0000-0001-8657-3448}
 \affiliation{
   \institution{University of Bristol}
   \city{Bristol}
   \country{United Kingdom}}
\email{ash.andrews@bristol.ac.uk}

\author{George Oikonomou}
\orcid{0000-0002-1684-6989}
 \affiliation{
   \institution{University of Bristol}
   \city{Bristol}
   \country{United Kingdom}}
 \email{g.oikonomou@bristol.ac.uk}

\author{Simon Armour}
\orcid{0009-0008-6319-9789}
 \affiliation{
   \institution{University of Bristol}
   \city{Bristol}
   \country{United Kingdom}}
 \email{simon.armour@bristol.ac.uk}

\author{Paul Thomas}
\orcid{0009-0003-5538-0338}
 \affiliation{
   \institution{University of Bristol}
   \city{Bristol}
   \country{United Kingdom}}
 \email{paul.thomas@bristol.ac.uk}

\author{Thomas Cattermole}
\orcid{0009-0001-2069-4922}
 \affiliation{
  \institution{Tiberos Research}
   \city{London}
   \country{United Kingdom}}
  \email{thomas.cattermole@tiberos.co.uk}

\renewcommand{\shortauthors}{Andrews et al.}

\begin{abstract}
As the Internet of Things (IoT) becomes more embedded within our daily lives, there is growing concern about the risk `smart' devices pose to network security. To address this, one avenue of research has focused on automated IoT device identification. This research is broadly motivated by the idea that the more we can know about our devices, the more secure the networks they are on can be. Research has however largely neglected the identification of IoT device firmware versions. There is strong evidence that IoT security relies on devices being on the latest version patched for known vulnerabilities. Identifying when a device has updated (has changed version) or not (is on a stable version) is therefore useful for IoT security. Version identification involves challenges beyond those for identifying the model, type, and manufacturer of IoT devices. Most obviously, the differences between versions are more subtle and therefore harder to detect. Moreover, because there has been relatively little research in this area, there are no widely available datasets that track devices' version changes over time. Consequently, traditional machine learning algorithms are ill-suited for effective version identification due to being limited by the availability of data for training. In this paper, we introduce an effective technique for identifying IoT device versions based on transfer learning. This technique relies on the idea that we can use a Twin Neural Network (TNN) --- trained at distinguishing devices --- to detect differences between a device on different versions. This facilitates real-world implementation by requiring relatively little training data. In more detail, we extract statistical features from on-wire packet flows, convert these features into greyscale images, pass these images into a TNN to output similarity scores, and determine version changes based on the Hedges' g effect size of the similarity scores. This allows us to detect the subtle changes present in on-wire traffic when a device changes version. To evaluate our technique, we set up a lab containing 12 IoT devices and recorded their on-wire packet captures for 11 days across multiple firmware versions. For testing data held out from training, our best performing model is shown to be 95.83\% and 84.38\% accurate at identifying stable versions and version changes respectively.
\end{abstract}



\received{10 July 2024}

\maketitle

\section{Introduction}
\label{sec:introduction}

The impact of the Internet of Things (IoT) is increasingly being felt across all aspects of society: in our homes (so-called `smart homes'), businesses (industrial IoT), and the public sector (smart cities). As more IoT devices become embedded within our daily lives, there is a growing fear about the potential risk they pose to security (not to mention privacy and safety). These fears have been realised with ``smart doorbells being an `easy target for hackers'"~\cite{BBC2} and seemingly innocuous children's toys allowing attackers access into networks and subsequently homes~\cite{BBC1}. To combat this risk, a body of research is developing devoted to automatically identifying IoT devices and understanding their behaviour. Recent work has largely focused on using supervised Machine Learning (ML) to classify wired (and wireless) network traffic in terms of device models, types and manufacturers. Underlying this work is the broad motivation that the more we can know about our IoT devices automatically, the more we can secure the IoT. More specific motivations include: automatically detecting when a rogue device enters a network and automatically detecting when a known device is behaving anomalously and therefore might be compromised. 

To fully identify an IoT device, there are four properties as defined by Meidan et al~\cite{Meidan2020}. These are: 1)~Type, 2)~Manufacturer, 3)~Model Number and 4)~Firmware Version. IoT device identification down to a software/firmware version level, specifically ensuring a device is running their latest version patched for known vulnerabilities~\cite{AndrewsWFIOT}, is a motivation that has been largely neglected~\cite{AndrewsCPSIOTSEC,AndrewsFMEC}. In order to be competitive, device manufacturers need to continuously release new products with short time-to-market and lowest possible prices. To this end, in this specific market, it is extremely hard to produce devices with strong security guarantees~\cite{Catuogno2023}
 and therefore flaws and vulnerabilities in devices are prevalent. A recent case with a flaw in Hikvision cameras allowed remote access~\cite{hikvis}, showing the importance of applying firmware updates to keep devices secure. However even with a serious vulnerability and a patch available, a whitepaper published 12 months later by Cyfirma~\cite{cyfirma} showed the patches had not been applied to over 80,000 internet reachable devices. This neglected motivation therefore has a concrete real-world application for IoT security: if we can automate version identification we can keep IoT devices up to date without user intervention, reducing their burden. 
 
 Version identification is however a more difficult problem to solve than other types of identification, such as device type or model. This is because the on-wire changes between versions are expected to be subtle and datasets are not publicly available containing version change information to train ML models. A method needs to be developed that can overcome these two challenges. This paper's contributions include:
 

\begin{itemize}
    \item A technique using transfer learning to identify device version changes without a need to train on a large amount of data, or having seen the data previously. More specifically using the output similarity scores from Twin Neural Networks with Hedges' g to identify both stable device versions and subtle device version changes,

    \item Obtaining a best accuracy of 95.83\% identifying stable versions and 84.38\% identifying version changes on a real world test-bed when using Hedges' g to identify subtleties on the Twin Neural Network output. The use of Hedges' g gives an improvement of approximately 20\% over alternative metrics, 

    \item A analysis of the difficulties in identifying IoT device firmware version changes seen in our testbed, specifically the subtle differences in on-wire signatures or in some cases no changes at all.
\end{itemize}

Our structure for the paper is as follows. In Section~\ref{sec:Motivation} we give a clear real-world motivation for how automated version identification can improve IoT security. We also describe the challenges of version identification and outline a method for addressing these challenges. In Section~\ref{sec:background} we examine the literature around IoT device identification and twin neural networks, including how we can interpret the output similarity scores when applied to IoT device versions. in Section~\ref{sec:method} we then use this knowledge for our proposed method, extracting our chosen features to generate pairs to train and then test the TNN on two scenarios - identification of devices with a stable version and identification of device version changes - using four measures of accuracy. In Section~\ref{sec:results} we briefly state the results from training the TNN and for our two test cases. We then complete a more in-depth analysis of the results with limitations of our approach, including analysis of which devices do and do not perform well. We state how these results relate to real-world use in Section~\ref{sec:discussion}. In Section~\ref{sec:conclusions} we conclude, outlining how our proposal can lead to future work in IoT device identification.

\section{Motivation}
\label{sec:Motivation}
\begin{figure}[h]
    \centering
    \includegraphics[width=0.75\textwidth]{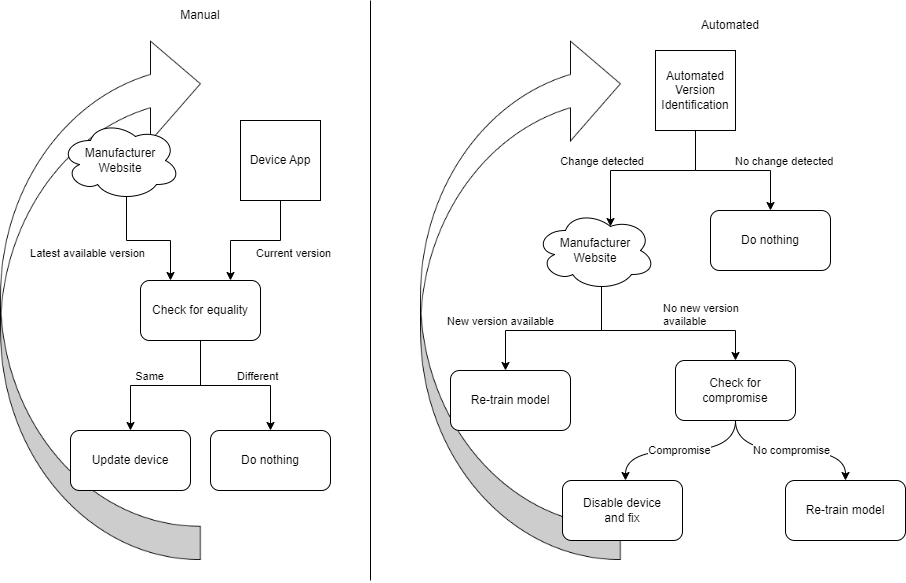}
    \caption{Comparison between manual and automated management of device versions in the IoT.}
    \Description[manual and automated management of iot devices are quite different.]{Diagram comparing manual on the left hand side and automated management on the right hand side of device versions.}
    \label{fig:manual_vs_automatic}
\end{figure}

To understand precisely how automatic version identification can help reduce the security burden in the IoT, it is worth considering what is demanded of users without automation. Without automation, a user must perform two preliminary administrative tasks for each IoT device to ensure it is up to date (see left side of Fig.~\ref{fig:manual_vs_automatic}):
\begin{enumerate}
    \item Find the latest version available for their IoT device.
    \item Find the current versions that their IoT device is using.
\end{enumerate}

The user is then tasked with comparing the results of (1) and (2): if the current version is not the latest version available, the user should update the device; if they are the same, the user should do nothing. Task (1) typically involves navigating to the relevant manufacturer website for the IoT device in question and checking the latest release notes. Task (2) typically involves accessing the IoT device's designated app and looking at the device’s details. Both of these tasks are time-consuming, and it is impractical to heavily rely on manual operations when securing a large number of IoT devices~\cite{zheng}. (2) is typically more time consuming than (1) because (1) can be done somewhat centrally (e.g. within one browser), while (2) may require accessing and navigating many different apps. Moreover, because navigating a website is more programmatic (e.g. via web-scraping) than accessing and navigating an app, (1) is currently more amenable to automation. Automated version identification (see right side of Fig.~\ref{fig:manual_vs_automatic}) removes the need to check the device app and requires checking the manufacturer website only when a device changes its behaviour. Automated version identification also makes it possible to detect compromises, when a device has not updated but is behaving anomalously. Both manual and automated approaches are cyclical throughout the lifetime of the device.

Our proposal rests on the idea that automated version identification can reduce the security burden on users by minimising the time they spend performing (2). In particular, by using ML to continuously detect changes in device behaviour, users do not need to manually access their devices' apps. This is a significant reduction in burden for the user. The security burden can therefore be front-loaded to the device, and at a moment when the administrative burden of the user is also present. Just as a user wants to set up their device only once, we believe that a user should set up their device in terms of security only once. Our solution provides that. Additionally, automated version identification can help users to detect compromises, because when a device is up to date but is behaving anomalously, a system could alert the user that the device should be checked manually. 

Automated IoT version identification however is a hard problem to solve in practice. IoT device identification has mainly been looked at as a classification problem where a technique involves training an ML model on known devices and seeing if it is possible to classify other devices of the same class. There are a few issues with this approach however. Firstly a classifier is only as good as the dataset used and although public datasets exist for use with ML techniques (\cite{Charyyev2021, Meidan2017, Perdisci2020a, Sivanathan2019a}) these datasets do not document initial version information or version changes over time. As devices change versions, the on-wire fingerprint can change and therefore the initial fingerprint may not still be relevant, as shown by Sivanathan et al~\cite{Sivanathan2020a}. Second, even with a dataset available for a ML model, there are a plethora of different devices and versions and so it is not feasible to be able to train a classifier on all possible combinations of data. To overcome the data problem, we can use transfer learning on an initial dataset with a ML approach that outputs similarity scores, such as a twin neural network. This allows us to re-use a pre-trained ML model on a future problem. A similarity based approach has an advantage over a classifier - classification algorithms seem ill-suited to the problem of devices running different versions, as typically a device's fingerprint will remain closer to its original classification than it will to other devices in the model's training set and be erroneously classified. A similarity approach however can state when a device no longer looks similar within a threshold, indicating something has changed.

\section{Background and Previous Work}
\label{sec:background}

\subsection{IoT Device Identification and Flow-based Features}
IoT device identification has been the focus of many studies with various different techniques employed. Techniques include: 1) Rule-based approaches: Creating rules based on a previous understanding of how a device behaves, 2) Unsupervised ML approaches: Using ML techniques to try and find patterns within the data with no prior training phase, 3) Supervised ML: Training and testing an ML model on known labelled data. For any technique the input data is network packets which can be obtained from either: 1) Active methods: Probing a device to elicit a response, or 2) Passive methods: Listening for packets that a device sends without external stimulation. In our study we only consider passive methods, as active methods could interfere with a device and cause it to malfunction. From this data, a common class of features known as `flow-based features' can be created which have been employed in various studies~\cite{Ammar2020, Msadek2019, Meidan2017, Sivanathan2019a, Sivanathan2020a, Wang2022} and therefore is a tangible research angle to pursue. A ﬂow is a succession of packets between a source and a destination characterized by the 4-tuplet (source IP, destination IP, source port, destination port)~\cite{Ammar2020, Agrafiotis2022}. These flows can have statistics calculated on them such as the maximum, minimum and average packet size. As well as packet size statistics, packet timing statistics can also be calculated. Some examples include the maximum, minimum and average time between packets that a device sends. Flows can be calculated across different protocols including NTP~\cite{Sivanathan2019a, Sivanathan2020a, Ammar2020}, DNS~\cite{Sivanathan2020a}, discovery protocols such as SSDP or mDNS~\cite{Ammar2020, Marchal2019}, and also encrypted protocols such as TLS~\cite{Msadek2019}. From analysing flows per protocols for each device, Sivanathan et al.~\cite{Sivanathan2020a} showed that there are ``signalling patterns" for protocols meaning devices do the same thing periodically, across different time periods~\cite{Marchal2019}, and this periodic communication is different per device~\cite{Perdisci2020a}.

\subsection{Image-based Classification and Twin Neural Networks}
Flow-based features across protocols can be fed into ML algorithms in various ways, including feeding in as a single feature vector, or transforming them first. One common transformation is to convert flow-based features into greyscale images. These techniques work by taking the chosen feature values and parsing them so that they can then be used as gradients of colour values between black (0) and white (255). Although not using the statistics but using the protocol payload data per flow for a varying number of packets, Lim et al.~\cite{Lim2019} and Xue et al.~\cite{Xue} generate greyscale images from network protocols (as opposed to network data from devices) in an attempt to classify the protocols, similar to Pathmaperuma et al.~\cite{Pathmaperuma2022} who use the same technique for user activity detection. Kotak et al.~\cite{Kotak2020} has a similar approach, but applied to IoT device identification. Converting network traffic to greyscale images also has uses within intrusion detection for identifying devices acting maliciously~\cite{Golubev2022, Agrafiotis2022, Taheri2018} or being part of a botnet~\cite{Hussain2021}.

These generated images can then be fed into various neural networks. Neural networks have shown to have use for network security including IoT device identification~\cite{Thompson2021} and intrusion detection~\cite{Kabir2021}. One such neural network is a twin neural network. Introduced by Bromley et al. \cite{bromley}, A Twin (also previously referred to as Siamese) Neural Network, henceforth referred to as a TNN, consists of two identical sub-networks (same architecture, parameters and weights) joined at their outputs. Each sub-network outputs an encoding as a vector, and the two vectors are then compared using a distance metric such as Euclidean distance. This output is then fed into a sigmoid function, such as contrastive loss, which allows a similarity score to be computed, stating how similar the two inputs are. The two metrics that a TNN outputs are accuracy and loss. Accuracy is how well the TNN is able to correctly determine if two input images are the same or not, and loss is a measure of how confident the model is about a prediction. If the images are determined to be the same, the TNN should output a value close to 0 and if they are different the TNN should output a value close to 1, with the threshold being 0.5. TNNs are trained on positive and negative pairs. Positive pairs are pairs that belong to the same class and negative pairs are pairs that belong to different classes. One of the main benefits of using a TNN is that it only requires a small amount of training data to produce accurate results and is also robust to class imbalance. TNNs have also been proposed as a method to solve the ML model retraining problem applied to IoT device identification~\cite{Trad2023}.

Although TNNs are mainly used for image recognition~\cite{snn_koch} or signature verification~\cite{Dey2017} they can also be used for network security including intrusion detection~\cite{Bedi2020}. Hindy et al.~\cite{Hindy2021} propose a system to detect network attacks using TNNs, with accuracy across test datasets between 84 and 91\%. The main benefits stated by the authors are that the TNN can learn from few instances, lessening the burden of collecting large amounts of data and labelling it. The TNN also allows a single model to be trained which in theory should then be able to classify new, unseen attacks. The main limitations of TNNs is that there needs to be an understanding of how to generate the most useful pairs to train the model - generating every possible combination is not feasible. There also needs to be an architecture for the TNN that is suitable for any future input data. 

\subsection{Statistical Analysis of TNN Output Similarity Scores Using Hedges' g}
Similarity score comparison for IoT device identification and determining the threshold for this comparison is an area that needs careful thought. Charyyev et al.~\cite{Charyyev2021} use similarity scores with locality-sensitive hashes per device, stating ``One may employ a device-specific threshold based on the variance of the signatures of the device''. One approach that can be used on a per device basis instead of a fixed threshold is Hedges' g. Hedges' g is a measure of the difference in means between datasets, and takes into account variability in datasets using the pooled standard deviation~\cite{Wilson2017}. It is defined as:

\[ \text{Hedges' g} = \frac{{{\overline{x}_1} - {\overline{x}_2}}}{{{s_p}}} \cdot J \text{ where } {s_p} = {{\sqrt{\frac{{(n_1 - 1) \cdot S_1^2 + (n_2 - 1) \cdot S_2^2}}{{n_1 + n_2 - 2}}}}} \text{ and where } J = \left(1 - \frac{3}{{4(n_1 + n_2) - 9)}}\right)\]

And where:
\begin{align*}
\phantom{=}{}& \overline{x}_1 \text{ and } \overline{x}_2 \text{ are the means of groups 1 and 2 respectively}    \\
\phantom{=}{}& {n_1} \text{ and } {n_2} \text{ are the number of samples in groups 1 and 2 respectively}  \\
\phantom{=}{}& {S_1} \text{ and } {S_2} \text{ are the standard deviations of groups 1 and 2 respectively} \\
\phantom{=}{}& {s_p} \text{ is the pooled standard deviation}   \\
\phantom{=}{}& {J}   \text{ is the bias correction factor}\\
\end{align*}
What the use of Hedges' g allows in the context of this paper is to remove the need to apply a fixed similarity threshold for all devices, instead indicating if a significant difference is detected between sample scores for a device. This facilitates generality to different sets of of devices. The Hedges' g score is interpreted in the following way:
\begin{itemize}
    \item{A positive Hedges' g value indicates that the first group has a larger mean than the second group.}
    \item{A negative Hedges' g value suggests that the second group has a larger mean than the first group.}
    \item{The magnitude of the Hedges' g value provides information about the effect size. Generally, a larger absolute value of g indicates a stronger effect.}
\end{itemize}

When assessing the practical significance of the effect size (\textit{g}) to decide if it is significant, the following values have been defined by Sawilowsky et al.~\cite{Sawilowsky2009}:
\begin{itemize}
    \item{$|g| \approx 0.01$ - Very Small effect}
    \item{$|g| \approx 0.2$  - Small effect}
    \item{$|g| \approx 0.5$  - Medium effect}
    \item{$|g| \approx 0.8$  - Large effect}
    \item{$|g| \approx 1.2$  - Very large effect}
\end{itemize}

It is shown that when a device changes version that the change is normally subtle and harder for a human to notice~\cite{AndrewsCPSIOTSEC} and so a `Medium effect' is likely to be observed. The differences are much larger~\cite{AndrewsCPSIOTSEC} between different devices and so it is likely that a `Large effect' is observed.

\subsection{Application to Our Study}

Flow-based features across multiple protocols from IoT network traffic have been shown in this section to be used extensively for IoT device identification, albeit for model identification, not version identification. These features are shown to have periodicity per protocol, and are also shown to be able to be converted into greyscale images suitable for use with neural networks, particularity TNNs, that output a similarity score between two inputs. This shows that generating greyscale images from network traffic and feeding into a TNN is a feasible approach for IoT device version identification, and using Hedges' g will allow a technique to identify subtle differences between device versions. We can use this to develop a technique for the following two scenarios applicable to the real world:

\begin{description}
    \label{list:scenarios}
    \item[Scenario 1:] Device is on the same version. We would expect the mean similarity score for subsequent days to be within a small threshold from the reference day (but not identical).
    \item[Scenario 2:] Device has updated, therefore there is a subtle change and the Hedges' g score will show a medium effect.
\end{description}

\begin{figure}[H]
    \centering    
    \includegraphics[width=0.6\linewidth]{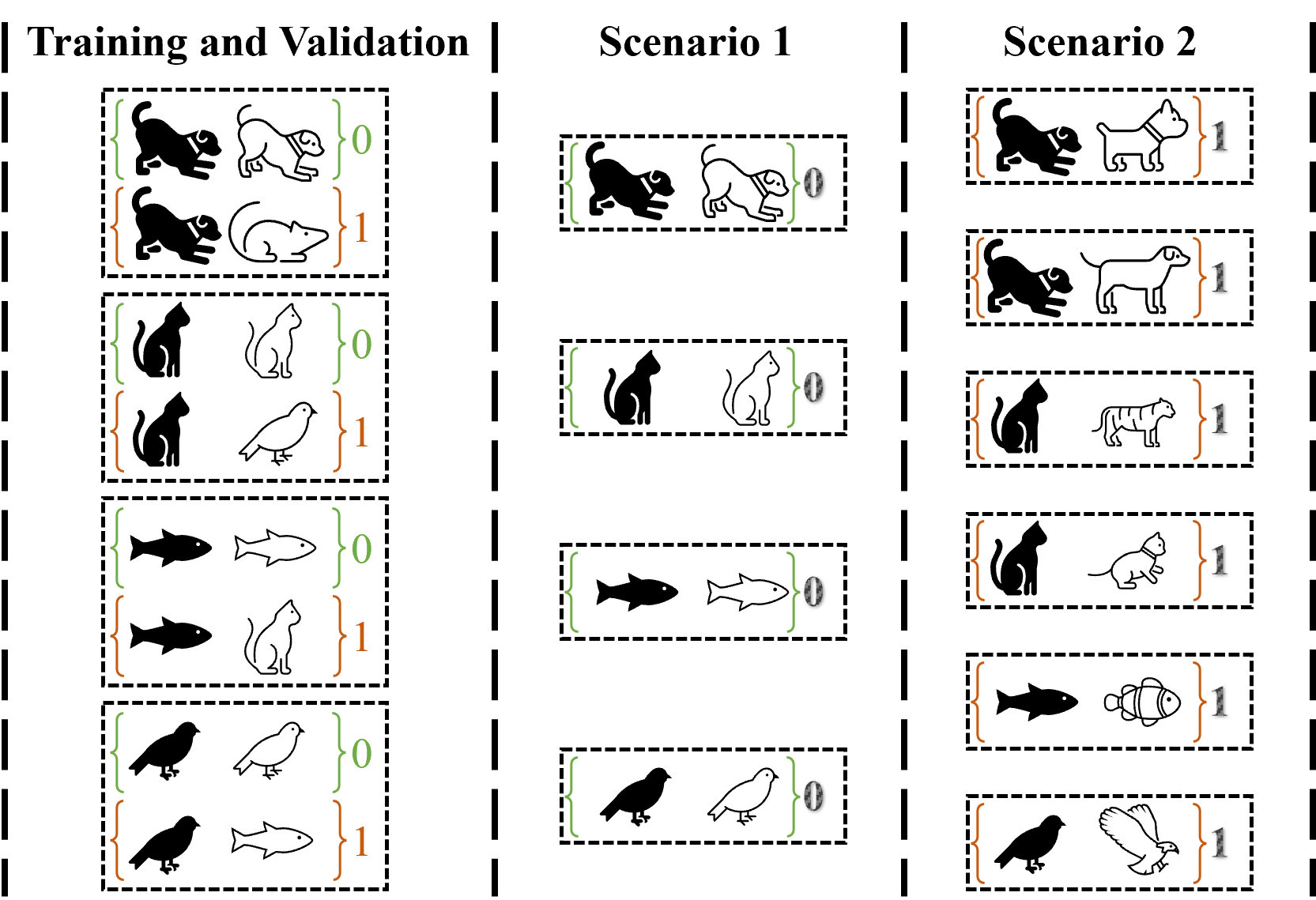}
    \caption{A visual analogy of our proposed technique as applied to animal classification.}
    \Description[Analogy of transfer learning with animals.]{A visual analogy of transfer learning applied to animals to relate to our technique. Training and validation is completed on the same animals, scenario 1 uses the same animals with different colours and scenario 2 uses different breeds of animals.}
     \label{fig:analogy}
\end{figure}

To understand how this works in practice, we can consider our technique for device version similarity analogous to animal breed similarity (shown in Fig.~\ref{fig:analogy}):
\begin{itemize}
\item A TNN is trained and validated on similar and dissimilar pairs. In our case, we have different devices each on a single version. This is analogous to different animals each of a single breed.

\item For Scenario 1 we want to ensure that our trained TNN can correctly identify the same device on the same version, albeit with slight on-wire differences. This is analogous to being able to identify the same animal and breed but with, for example, a slightly different colour pattern.

\item For Scenario 2 we want to ensure that our trained TNN can pick up subtle changes to device versions. This is analogous to identifying subtle differences between different breeds of the same animal.

\item Both cases use transfer learning as mentioned in Section~\ref{sec:Motivation} - the trained TNN only knows about different devices each on a single version, not different versions per device. This is analogous to the trained TNN knowing about different animals each of a single breed and not different breeds per animal.

\end{itemize}

\section{Method}
\label{sec:method}

For our technique, we set ourselves two requirements, based on the Scenarios in Section~\ref{sec:background}, to be explored in this study:

\begin{description}
    \item[Requirement 1:] The technique should reliably identify a device running the same version and not erroneously state there is a difference when there is not,
    \item[Requirement 2:] The technique should be able to correctly identify when a device has changed version.
\end{description}

As mentioned in Section~\ref{sec:Motivation}, public datasets exist but they do not contain device version information over time, and so to satisfy requirements 1 and 2 we needed to set up our own IoT lab with a traffic capture system to capture all traffic from the IoT devices within the network over time, as they change firmware version. There are an average of 9 devices per UK household~\cite{UKParliament} and so we set up a lab to mimic a smart home setup with a range of 12 smart devices, connected to a network switch. Our lab devices all generate network traffic and this traffic is mirrored and captured at the switch, called the `traffic sniffer'. We use this traffic to train, validate and test our TNN for our two requirements. Our exact methodology around feature selection, image generation, TNN training and testing is covered in the following subsections. This traffic capture system is shown in the top left of Figures~\ref{fig:rough-arch} and ~\ref{fig:rough-arch2} that follow Section~\ref{sec:method_eval}.

\subsection{Traffic Analysis and Feature Selection}
As shown in Section~\ref{sec:background}, flow statistics for various protocols can be used for IoT device identification. Based on the literature in Section 3.1, for this study we consider the following protocols: TCP, HTTP, OCSP, TLS, UDP, DNS, NTP, mDNS, SSDP, SIP, QUIC, ICMP and IGMP. There are two types of traffic to use to generate flow statistics per protocol, with advantages and disadvantages to each:

\begin{enumerate}
 \item \textbf{\textit{Traffic originating from the device only going out of the network}} - This has the advantage of not including potential noise from other devices within the local network and is more likely to be applicable across networks. The disadvantage however is that this would not capture anything the device does internally to a network that could be unique, such as pinging a local gateway periodically or setting up a local mesh network.
 
 \item \textbf{\textit{Traffic only between devices locally}} - This has the advantage of giving an understanding of how devices interact with each other within a network. This may however only be a small amount of data, or the inter-device communications may not be applicable to other networks containing different device combinations.
 
\end{enumerate}

We use both (1) and (2), allowing us to use all traffic generated per device potentially giving us the richest data.

\begin{figure}[H]
    \centering
    \fboxsep=0pt
    \fbox{\includegraphics[width=0.98\linewidth]{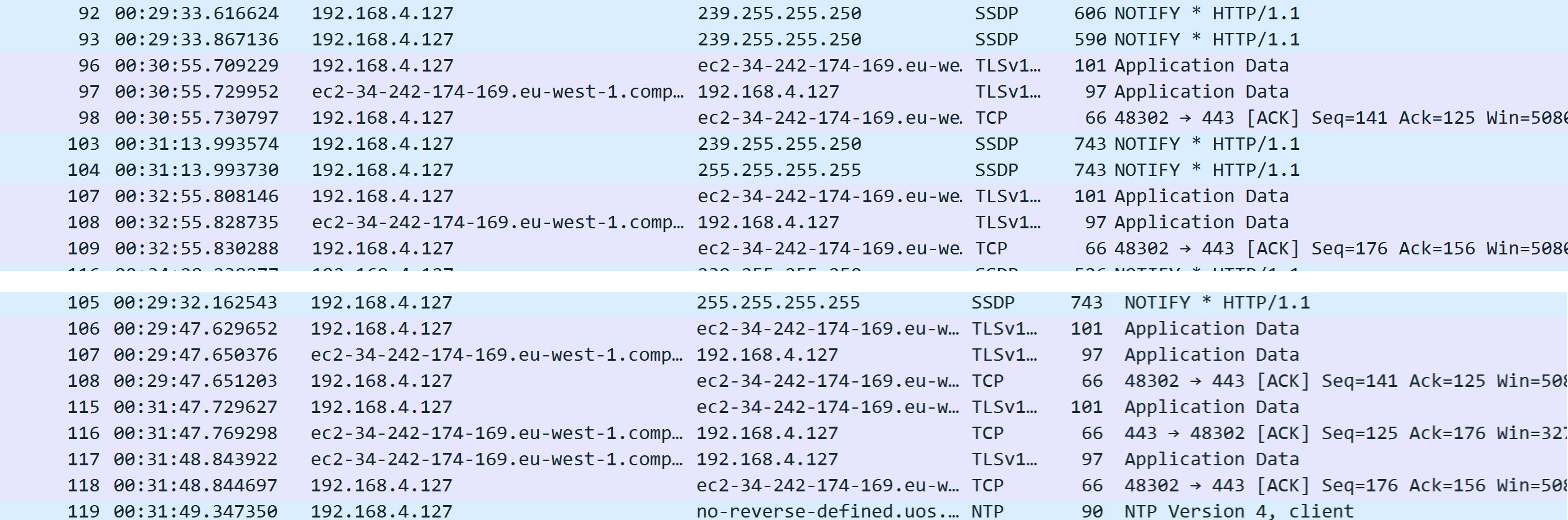}}
    \caption{Device packets either side of a time window boundary.}
    \Description[TCP and TLS exchanges for two days of traffic]{Example TCP and TLS traffic for a single device across two days, showing how there is a slight time shift.}
    \label{fig:sliding-bins-1}
\end{figure}

When calculating flow statistics for network traffic, there is a decision to make around 1) the time window to create all samples over, i.e. the time window and 2) the sample time to calculate our statistics over, i.e, the time period. For 1), we choose a time window of 6 hours between 00:00 and 06:00 for each device. Our reasoning is as follows: We firstly start our processing at the start of a new day for a device as you'd expect a device to do specific things daily. We would also expect the traffic during this time to be stable and not involve user interaction. 6 hours is also 25\% of the data available for a device and so there is a suitable trade-off here between amount of data and processing vs the benefit gained. For 2) we again have a trade-off of processing needs vs benefit. With a smaller sample size, we get more sample points but more processing needs. With a larger sample size, we may not pick up many flow features but less processing power is needed. We also need to consider our sample window boundaries. As shown in Fig.~\ref{fig:sliding-bins-1} we can see the device clearly has a pattern of 6 packets within the TLS traffic around a potential time window boundary of 00:30, occurring at 00:29:47 and 00:31:47. With a fixed window approach, these packets would be included in two separate sample bins, the first being 00:15:00 - 00:30:00 and the second being 00:30:00 - 00:45:00. We can see however in that same figure, possibly due to network delays or clock drift, these same packets a few days later have shifted, and from this all 6 packets would then appear within the sample bin between 00:30:00 - 00:45:00. Because of this we use a 20\% sliding window approach across the network PCAP to capture this temporal information which is sufficient, so in this example our sample bins would be 00:24:00 - 00:39:00, 00:36:00 - 00:51:00 capturing all of the information.

To summarise, we calculate our flow statistics over a time window between 00:00 and 06:00. Within this, we calculate flow statistics over 15 minute windows with a 20\% overlap resulting in flow periods 00:00 - 00:15, 00:12 - 00:27, 00:24 - 00:39 etc. giving us 30 datapoints per device per day.

\subsection{Conversion to Greyscale Image}

With our flow features extracted, we then create greyscale images per flow period (15 minute sliding window) over 6 hours for use with our TNN. For each device we therefore have a maximum of 30 images per day. There is difficulty in providing a technique that can capture all the information across protocols as one set of features, while still considering how the features can affect each other. This is why our implementation of a greyscale image can work well - it allows patterns both between and within protocols to be noticed. An example image is given in Fig.\ref{fig:sample-input-img}. We do the following to create our greyscale images. The Y axis per image is the 13 protocols we have calculated statistics over mentioned in the previous subsection. The X axis becomes the features calculated from the flow statistics, shown in Table~\ref{tab:features_used_in_image}. For each feature, we normalize the value between 0 and 255 to be suitable for a greyscale image, but group together features for normalization in the following way for 4 different sets of features (shown separated as red dashed lines within Fig.~\ref{fig:sample-input-img}):

\begin{figure}[h]
    \centering    
    \includegraphics[width=0.85\linewidth]{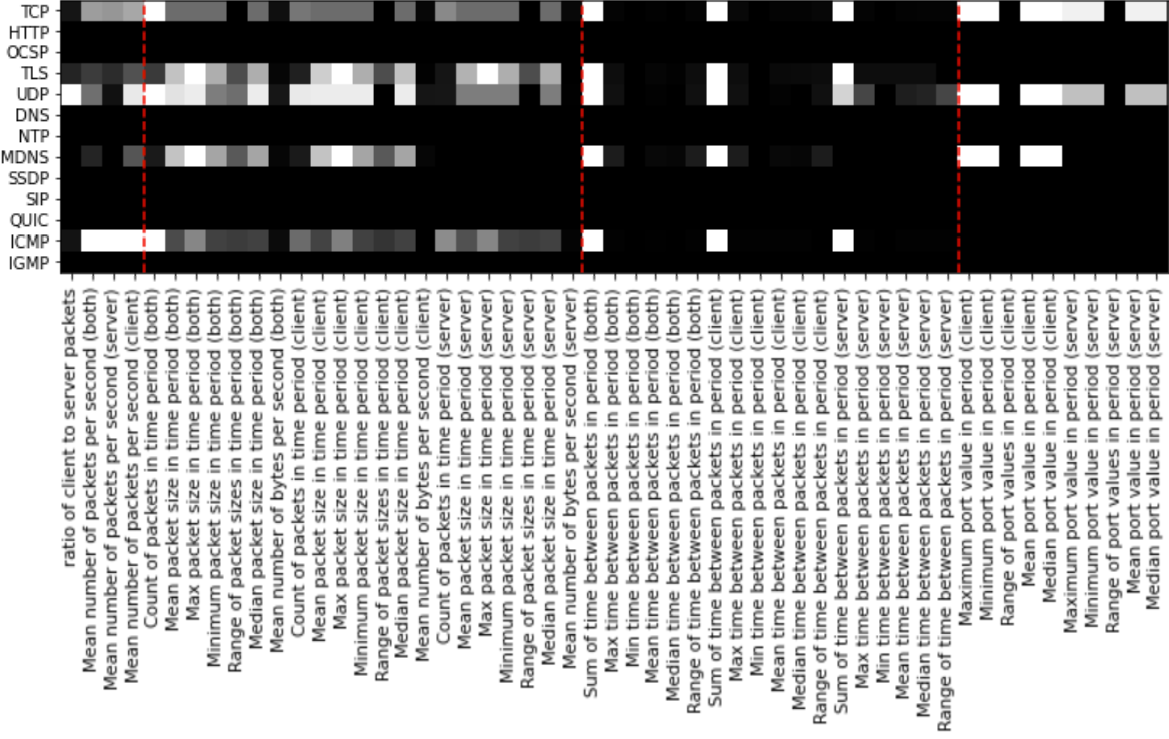}
    \caption{Example greyscale image, showing protocols on the Y axis and features on the X axis.}
    \Description[Sample input image]{Sample input greyscale image for the technique showing 9 protocols on the Y axis and 53 features on the X axis.}
    \label{fig:sample-input-img}
\end{figure}

\begin{itemize}
\item The first set are relevant across the protocols, and are normalized \textbf{\textit{across the protocols}},
\item The second set are related to packets and generally large values and are normalised \textbf{\textit{per protocol}},
\item The third set are related to timings and are generally small values and normalised \textbf{\textit{per protocol}},
\item The fourth set are related to ports and will have a standard maximum and minimum value and are normalised \textbf{\textit{per protocol}}.
\end{itemize}

This then gives an X by Y image per device per sample period with all values between 0 (black) and 255 (white). This gives a fingerprint for a device which can look unique at a high level, and may have more nuanced pixel differences that a TNN can identify. We clean our dataset of input images by removing any images that contain no useful data, i.e. are fully black or white. We can then visualise these images before passing into a TNN. We provide a visualisation with all information in Fig.\ref{fig:sample-input-img}, but omit the axis labels in subsequent sections for space saving reasons. Once the greyscale images are created per time period per device, we then need to make pairs to train the TNN.

\small
\begin{table}[H]
\begin{tabular}{|c|c|c|c|c|}

\multicolumn{1}{p{100pt}}{\textbf{Feature Set}} & 
\multicolumn{1}{p{84pt}}{\raggedright \textbf{For Client (C), Server (S) or Both (B)}} & 
\multicolumn{1}{p{145pt}}{\raggedright \textbf{Feature Description}} & 
\multicolumn{1}{p{55pt}}{\raggedright \textbf{Image Section Feature Count}}\\ 
\hline

\multicolumn{1}{p{100pt}}{\multirow [t]{2}{=}{Statistics across all}} & 
\multicolumn{1}{p{84pt}}{C | S} & 
\multicolumn{1}{p{145pt}}{\raggedright Ratio of client to server packets} & 
\multicolumn{1}{p{55pt}}{\multirow [t]{2}{=}{\centering 4}} \\ 
\multicolumn{1}{l}{protocols} & 
\multicolumn{1}{p{84pt}}{C | S | B} &
\multicolumn{1}{p{145pt}}{\raggedright Mean number of packets per second}  & \multicolumn{1}{l}{}\\ 
\hline

\multicolumn{1}{p{100pt}}{\raggedright Packet features across each} & 
\multicolumn{1}{p{84pt}}{\multirow [t]{7}{=}{\raggedright C | S | B}} & 
\multicolumn{1}{p{145pt}}{\raggedright Count of packets in time period} & 
\multicolumn{1}{p{55pt}}{\multirow [t]{7}{=}{\centering 21}}\\ 

\multicolumn{1}{p{100pt}}{protocol} & \multicolumn{1}{l}{} & 
\multicolumn{1}{p{145pt}}{\raggedright Mean packet size in time period}  & \multicolumn{1}{l}{}\\ 
\multicolumn{1}{p{100pt}}{} & \multicolumn{1}{l}{} & 
\multicolumn{1}{p{145pt}}{\raggedright Max packet size in time period} & \multicolumn{1}{l}{}\\ 
\multicolumn{1}{p{100pt}}{} & \multicolumn{1}{l}{} & 
\multicolumn{1}{p{145pt}}{\raggedright Minimum packet size in time period} & \multicolumn{1}{l}{}\\ 
\multicolumn{1}{p{100pt}}{} & \multicolumn{1}{l}{} & 
\multicolumn{1}{p{145pt}}{\raggedright Range of packet sizes in time period}  & \multicolumn{1}{l}{}\\ 
\multicolumn{1}{p{100pt}}{} & \multicolumn{1}{l}{} & 
\multicolumn{1}{p{145pt}}{\raggedright Median packet size in time period} & \multicolumn{1}{l}{}\\ 
\multicolumn{1}{p{100pt}}{} & \multicolumn{1}{l}{} & 
\multicolumn{1}{p{145pt}}{\raggedright Mean number of bytes per second} & \multicolumn{1}{l}{}\\ 
\hline

\multicolumn{1}{p{100pt}}{\raggedright Timing features across each} & 
\multicolumn{1}{p{84pt}}{\multirow [t]{6}{=}{\raggedright C | S | B}} & 
\multicolumn{1}{p{145pt}}{\raggedright Sum of time between packets in period}  & 
\multicolumn{1}{p{55pt}}{\multirow [t]{6}{=}{\centering 18}}\\ 

\multicolumn{1}{p{100pt}}{protocol} & \multicolumn{1}{l}{} & 
\multicolumn{1}{p{145pt}}{\raggedright Max time between packets in period}  & \multicolumn{1}{l}{}\\ 
\multicolumn{1}{p{100pt}}{} & \multicolumn{1}{l}{} & 
\multicolumn{1}{p{145pt}}{\raggedright Min time between packets in period} & \multicolumn{1}{l}{}\\ 
\multicolumn{1}{p{100pt}}{} & \multicolumn{1}{l}{} & 
\multicolumn{1}{p{145pt}}{\raggedright Mean time between packets in period} &  \multicolumn{1}{l}{}\\ 
\multicolumn{1}{p{100pt}}{} & \multicolumn{1}{l}{} & 
\multicolumn{1}{p{145pt}}{\raggedright Median time between packets in period} &  \multicolumn{1}{l}{}\\ 
\multicolumn{1}{p{100pt}}{} & \multicolumn{1}{l}{} & 
\multicolumn{1}{p{148pt}}{\raggedright Range of time between packets in period} &  \multicolumn{1}{l}{}\\ 
\hline

\multicolumn{1}{p{100pt}}{\raggedright Port features across each} & 
\multicolumn{1}{p{84pt}}{\multirow [t]{5}{=}{\raggedright C | S}} & 
\multicolumn{1}{p{145pt}}{\raggedright Maximum port value in period}  & 
\multicolumn{1}{p{55pt}}{\multirow [t]{5}{=}{\centering 10}}\\ 

\multicolumn{1}{p{100pt}}{protocol} & \multicolumn{1}{l}{} & 
\multicolumn{1}{p{145pt}}{\raggedright Minimum port value in period}  & \multicolumn{1}{l}{}\\ 
\multicolumn{1}{p{100pt}}{} & \multicolumn{1}{l}{} & 
\multicolumn{1}{p{145pt}}{\raggedright Range of port values in period} & \multicolumn{1}{l}{}\\ 
\multicolumn{1}{p{100pt}}{} & \multicolumn{1}{l}{} & 
\multicolumn{1}{p{145pt}}{\raggedright Mean port value in period} & \multicolumn{1}{l}{}\\ 
\multicolumn{1}{p{100pt}}{} & \multicolumn{1}{l}{} & 
\multicolumn{1}{p{145pt}}{\raggedright Median port value in period} & \multicolumn{1}{l}{}\\ 
\hline

\end{tabular}
\caption{Features used in greyscale image creation and which section of the image they occupy.}
\label{tab:features_used_in_image}
\end{table}
\normalsize

\subsection{Training the Twin Neural Network}

\begin{figure}[H]
    \centering
    \includegraphics[width=1\linewidth]{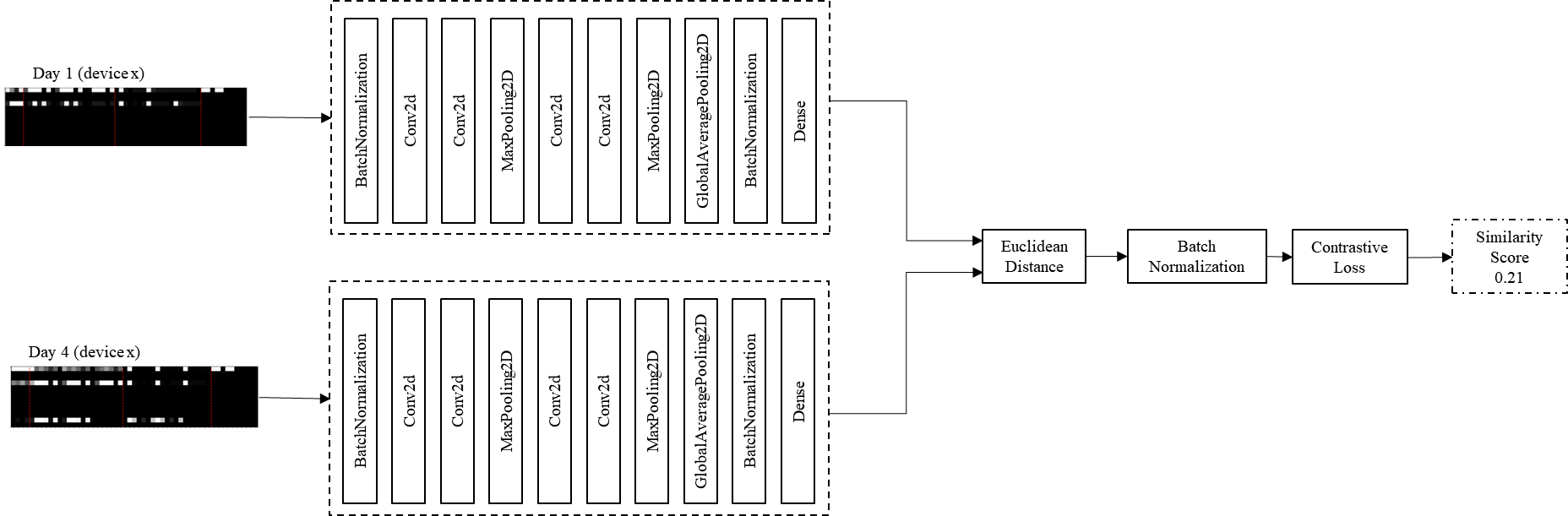}
    \caption{Internal architecture of our TNN showing the chosen layers with contrastive loss.}
    \Description[Architecture of the TNN.]{Internal architecture of our TNN showing the 10 chosen layers with batch normalisation and contrastive loss.}
    \label{fig:siamese-diagram}
\end{figure} 

\begin{table}[H]
\begin{tabular}{|c|c|c|c|c|}
\hline
Number of Classes & {Batch Size} & {Epochs} & {Early Stopping} & {Margin for Contrastive loss} \\
\hline
12 & 50          & 50   & Yes - Validation Loss & 1 \\ \hline

\end{tabular}
    \caption{Parameters chosen to create our TNN.}
    \label{tab:siamese-params}
\end{table}

\begin{figure}[H]
    \centering
    \includegraphics[width=1\linewidth]{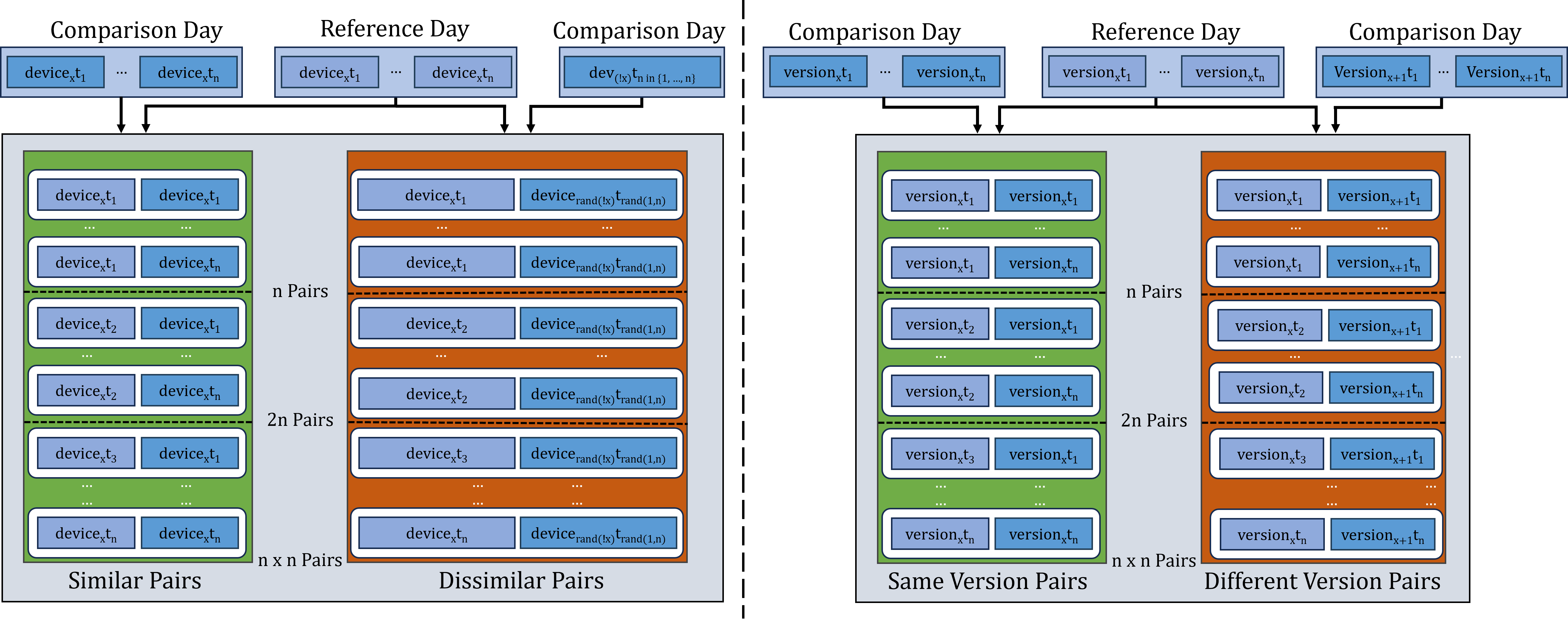}
    \caption{The creation of pairs for training and validation (left side) and testing (right side) in diagram form. For our study, the number of devices is 12 and the pictorial flow is for all integer values of x between 1 and 12. We expect n, the maximum value for the index of the time window, to be 30 in most cases, if all timeslot images exist.}
    \Description[Diagram showing how pairs are created.]{The creation of pairs for training and validation (left side) and testing (right side) in diagram form. For our study, the number of devices is 12 and the pictorial flow is for all integer values of x between 1 and 12. We expect n, the maximum value for the index of the time window, to be 30 in most cases, if all timeslot images exist.}
    \label{fig:testing_pairs_explained}
\end{figure}

As we are using a \textit{twin} neural network, the architecture of our TNN contains two identical subnetworks, with each arm receiving an input. We feed in two images and get the TNN to produce a similarity score between the two inputs. The architecture with our chosen layers is shown in Fig.~\ref{fig:siamese-diagram}. Our parameters used when creating our TNN are shown in Table~\ref{tab:siamese-params}. We create positive pairs, pairs where both images belong to the same class, and negative pairs, where the images are from different classes. We ensure our input dataset is balanced, i.e. the number of positive and negative pairs are the same, and the negative pairs are varied across classes by generating pairs in the following way (as shown in Fig.~\ref{fig:testing_pairs_explained}):

\begin{enumerate}

\item \textbf{Generating similar pairs for training and validation:} We take each image from \textbf{day 1} and pair it to every image for the same device on a single version from \textbf{day 2 for training} or \textbf{day 3 for validation}.

\item \textbf{Generating dissimilar pairs for training and validation:} We again take each image from \textbf{day 1} and pair it with a random selected image from a randomly selected difference device on a single version from \textbf{day 2 for training} or \textbf{day 3 for validation}.

\item \textbf{Generating similar pairs for experiment 1:} For each device, we take each image from \textbf{day1} and pair it to every image for the same device on the same version from \textbf{days 4-7}.
\item \textbf{Generating dissimilar pairs for experiment 2:} For each device, we take each image from \textbf{day1} and pair it to every image for the same device on the a different version from \textbf{days 8-11}.

\end{enumerate}

Using this approach, the maximum number of similar pairs per device, per day is 30 * 30 = 900, and we therefore randomly select the same number of dissimilar pairs. This however is reliant on there being enough data from a device to generate statistics, i.e. some traffic in every 15 minute window. If not, the number of samples, and therefore pairs, will decrease. We generate the pairs in this way for two reasons. First, we use chronological days, which is preferred to a random split of data~\cite{AndrewsWFIOT} and second, we consider the real-world use of our tool being used within a network when training, validating and testing a TNN. A typical deployment would be as follows. Our tool would be deployed and collect network traffic from all devices. The data from day 1 is the baseline day, showing what the network should look like under normal circumstances. We then train and validate our TNN on the training (day 1 and 2) pairs and validation (day 1 and 3) pairs, which allows our TNN to learn what similar device samples look like. We then test on subsequent days paired with day 1.

\begin{figure}[ht]
     \centering
     \includegraphics[width=0.75\textwidth]{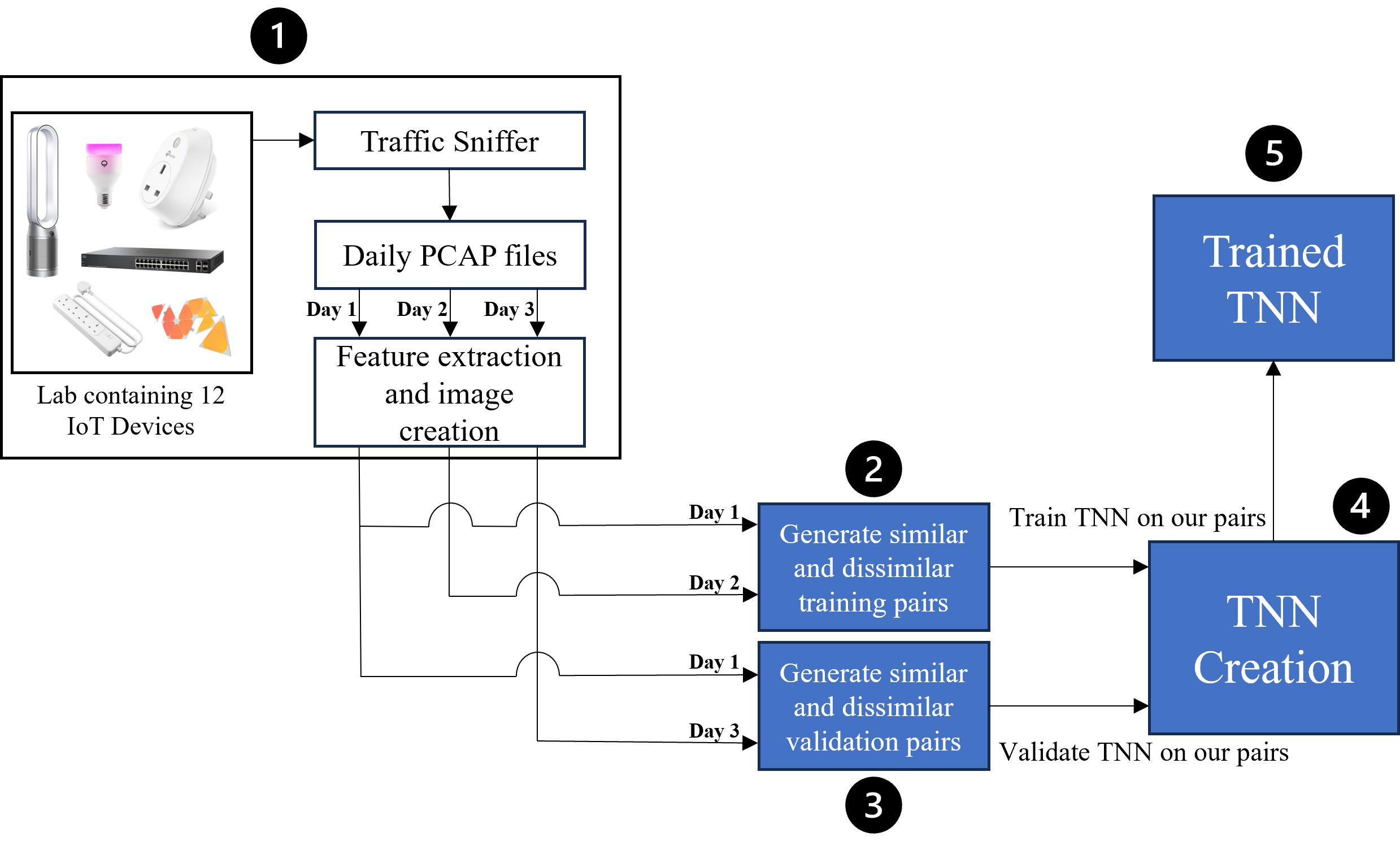}
     \caption{System architecture for creating our TNN.}
     \Description[System architecture.]{System architecture of the full system from input traffic capture, to generating pairs to creating and training a TNN.}
     \label{fig:rough-arch}
 \end{figure}

Our full end-to-end system for training our TNN is shown in Fig.~\ref{fig:rough-arch}. Our IoT lab traffic is sniffed to generate PCAP files per device, which then allows feature extraction and image creation (1). Day 1 and 2 data is used to generate the similar and dissimilar training pairs (2), with days 1 and 3 data used to generate the similar and dissimilar validation pairs (3). The TNN is then created using the training and validation pairs, with early stopping monitoring validation loss (4). This then gives us a Trained TNN (5).

\subsection{Evaluating the Technique}
\label{sec:method_eval}

To evaluate our technique, we will perform the following experiments linking back to our requirements 1 and 2 from Section~\ref{sec:method}. We will first train our TNN on devices where the versions do not change and ensure that across training and validation we get a good accuracy score and are not overfitting on our data. We will then perform the following two specific experiments using our trained TNN:

\begin{description}
    \item[Experiment 1:] We will test our TNN on multiple days of data from each device (where there are no version changes), outputting the overall accuracy per device. 
    \item[Experiment 2:] We will test our TNN on multiple days of data from each device where there is a version change. We will again output the overall accuracy per device.
\end{description}

For experiments 1 and 2 we do the following. We feed the pairs into the trained TNN, get the similarity scores for the inputs and use these to produce accuracy scores. For our technique we need a way to quantitatively measure the difference between the aggregated similarity scores for device pairs fed into the neural network between the reference and subsequent days. We consider 4 different indicators of accuracy:
\begin{enumerate}
    \item All Samples: The percentage across all days considering all samples where they correctly show as 'similar'.
    \item Majority Samples: The percentage of the days where the number of samples correctly shown as 'similar' is more than 'dissimilar' for each day.   
    \item Majority Mean Threshold: The percentage of the days where the mean of all the samples is above the chosen threshold value of 0.5.
    \item Hedges' g: The percentage of the days where the Hedges' g score of the mean of all the samples is above the chosen effect size value of 0.5 and positive.
\end{enumerate}

We use Hedges' g due to the fact it doesn't require the datasets to be normally distributed, which in our case holds true as we would expect our reference day values to all be close to either 0 or 1 if the TNN is working correctly. Because of not always having 900 samples, we then need to use Hedges' g as it accounts for biases introduced with small sample sizes and is therefore useful when comparing our datasets that may have a different number of samples (Hedges' g is generally useful when the number of samples is less than 50). When using our TNN, the expectation is that the device input images are dissimilar enough to be classified using the default similarity threshold of 0.5 for similar or dissimilar. This may hold true for different device models, but with device versions it is likely that the change will be more subtle and therefore unlikely that this would result in a large similarity score change, above the 0.5 threshold. This allows us to understand whether our technique is stable for experiment 1 and can detect changes for experiment 2, and how different measures of accuracy perform. As we vary the dissimilar images randomly, we will perform 10 runs, giving rise to 10 different created TNNs, to see how stable the results are.

\begin{figure}[h]
     \centering
     \includegraphics[width=1\textwidth]{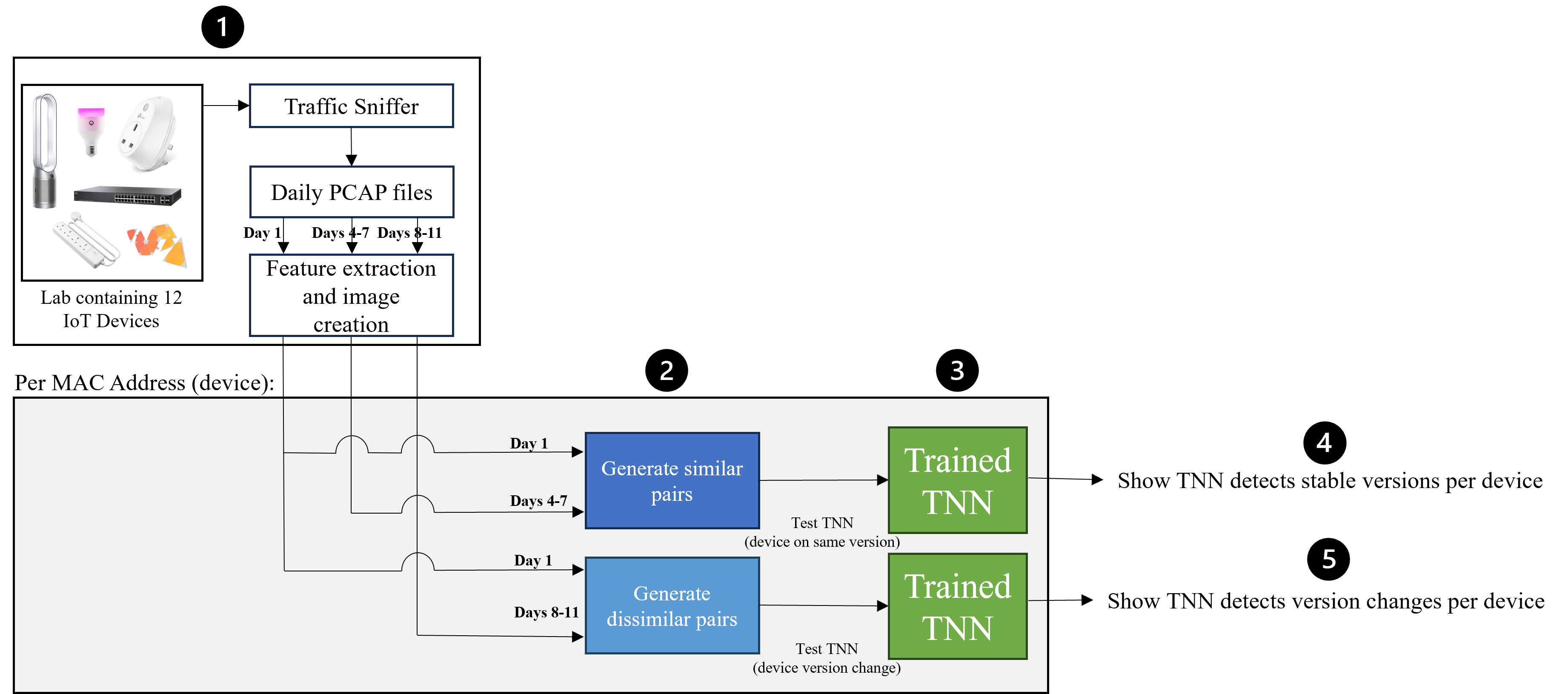}
     \caption{System architecture for testing the trained TNN with two experiments.}
     \Description[System architecture for training.]{System architecture for training starting with the input traffic capture into pair generation and feeding into the trained TNN, showing the TNN can detect stable versions and version changes.}
     
     \label{fig:rough-arch2}
 \end{figure}

Our full end-to-end system for testing our TNN is shown in Fig.~\ref{fig:rough-arch2}. Our IoT lab traffic is sniffed to generate PCAP files per device, which then allows feature extraction and image creation for 11 days (1). For every device, pairings for day 1 to days 4, 5 6 and 7 are generated as similar pairs (same version) and pairings for days 1 to days 8, 9, 10 and 11 are generated as dissimilar pairs (version change) (2). These are both passed in the previously trained TNN (3) with the intention to show that the TNN can detect stable versions per device (4) and version changes (5).

\section{Results}
\label{sec:results}

We first review the results of our two experiments at a high level, with a more in-depth discussion of the results in the section \ref{sec:discussion}. We will consider the results across our 10 runs generally.

\subsection{Training and Validating the TNNs}

\begin{figure}[h]
    \centering
    \includegraphics[width=0.90\linewidth]{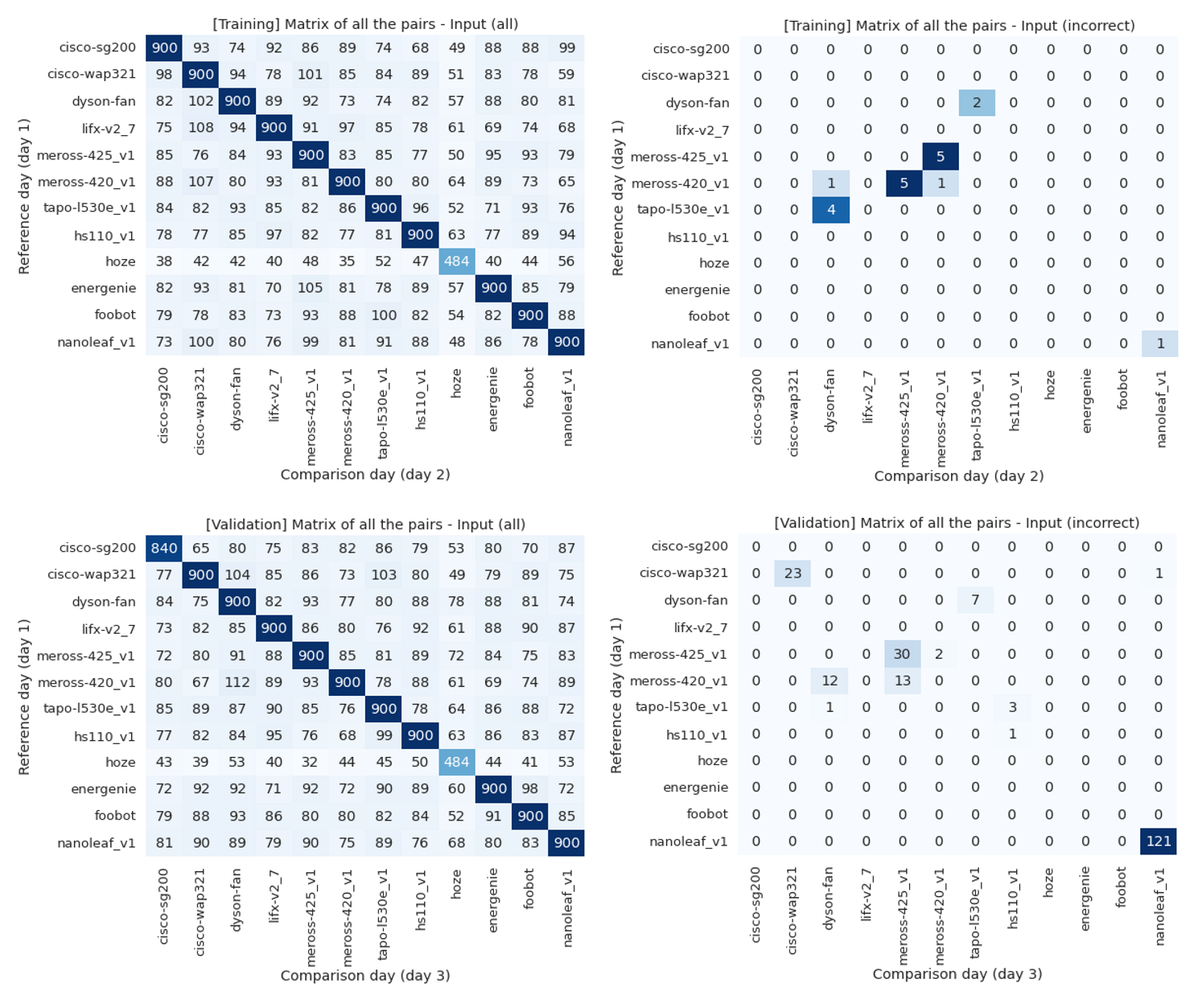}
    \caption{Training (top left) and validation (bottom left) image distribution for run 4, including mis-classifications for training (top right) and validation (bottom right).}
    \label{fig:training_pairs}
    \Description[input input distributions.]{4 input images showing distributions. Training (top left) and validation (bottom left) image distribution for run 4, including mis-classifications for training (top right) and validation (bottom right).}
\end{figure}

When training our TNNs, we have 1800 possible pairs for training -- 900 similar and 900 dissimilar -- for our time window size across our time period. For our training set, we found that this maximum number holds true for all except one device. Our Hoze device only had 484 samples, meaning only just over half of the timeslots had data. For our validation set, we found again that our Hoze device had 484 samples, but also our Cisco SG200 device had slightly less than 900 with 840 samples. This however is still enough to train the TNNs adequately as we ended up with 20768 samples in total to train our TNNs, and 20648 samples for validation. For each of the 10 models we train, the similar pairs are always the same and we vary the dissimilar images (although there are always a maximum of 900 of each per device). The distribution of our training pairs for our highest performing run, and also the main mis-classifications are shown in Fig.~\ref{fig:training_pairs}.

The overall results of training our TNNs are shown in Table~\ref{tab:TNN_training_results}. We can see that through 10 separate iterations of TNN training with varied dissimilar images the accuracy and loss scores for all samples (both similar and dissimilar images) are all at around 99\%. This shows our technique generally works and the TNNs can distinguish between similar and dissimilar pairs well. Note that these results are with the TNN using a default threshold of 0.5 for a similarity decision. For our specific use case, we are interested only in the similar pairs to see whether the technique works well enough to correctly identify similar images for devices. This is analogous to one device with one MAC address working as expected, with a stable version. Considering just the similar pairs again, the TNNs score highly at approximately 99\%. From these results we would expect our TNNs to perform well at identifying stable device versions when comparing subsequent days of unseen data to the day 1 baseline. We may also expect it to perform fairly well on completely unseen and different data (version change data), as it should pick up the nuances in version difference due to the small loss values. Based on the validation results in Fig.~\ref{fig:training_pairs}, we may expect the TNN to perform worst with the Nanoleaf Shapes device, with some other misclassifications occurring sporadically too.

\footnotesize
\begin{table}[H]
\centering

\begin{tabular}{l|c|c|c|c|c|c|c|c|c|c|l|}
\cline{2-12}
     &
  R1 &
  R2 &
  R3 &
  R4 &
  R5 &
  R6 &
  R7 &
  R8 &
  R9 &
  R10 &
  \multicolumn{1}{c|}{Average} \\ \hline
\multicolumn{1}{|l|}{Training Loss (All Samples)} &
  0.0020 &
  0.0025 &
  0.0020 &
  0.0015 &
  0.0028 &
  {\textbf{0.0008}} &
  0.0010 &
  0.0010 &
  0.0029 &
  0.0026 &
  0.0019 \\ \hline
\multicolumn{1}{|l|}{Validation Loss (All Samples)} &
  0.0096 &
  0.0101 &
  0.0095 &
  0.0093 &
  0.0108 &
  0.0103 &
  0.0065 &
  {\textbf{0.0059}} &
  0.0103 &
  0.0091 &
  0.0091 \\ \hline
\multicolumn{1}{|l|}{Training Accuracy (All   Samples)} &
  0.9979 &
  0.9970 &
  0.9979 &
  0.9982 &
  0.9966 &
  {\textbf{0.9989}} &
  0.9987 &
  0.9988 &
  0.9964 &
  0.9968 &
  0.9977 \\ \hline
\multicolumn{1}{|l|}{Training Accuracy (Similar   Only)} &
  0.9970 &
  0.9937 &
  0.9979 &
  0.9994 &
  0.9966 &
  0.9979 &
  {\textbf{1.0000}} &
  0.9999 &
  0.9988 &
  0.9998 &
  0.9981 \\ \hline
\multicolumn{1}{|l|}{Validation Accuracy (All   Samples)} &
  0.9885 &
  0.9881 &
  0.9898 &
  0.9896 &
  0.9878 &
  0.9893 &
  0.9932 &
  {\textbf{0.9937}} &
  0.9880 &
  0.9896 &
  0.9898 \\ \hline
\multicolumn{1}{|l|}{Validation Accuracy (Similar   Only)} &
  0.9774 &
  0.9800 &
  0.9841 &
  0.9809 &
  0.9827 &
  0.9801 &
  0.9873 &
  {\textbf{0.9891}} &
  0.9787 &
  0.9830 &
  0.9823 \\ \hline
\end{tabular}
\caption{TNN training and validation accuracy and loss for each of the 10 runs, to 4 decimal places. The run with the highest result per metric is shown in bold.}
\label{tab:TNN_training_results}
\end{table}
\normalsize

\subsection{Stable Version Test}

In the first experiment we test whether our trained TNN can successfully identify stable device versions. The input pairings are day 1 paired with days 4, 5, 6 and 7, which are unseen data but for known classes. Our overall accuracy results are shown in Table~\ref{tab:TNN_model_results}, and we now go through each of our accuracy metrics in turn. We see when considering all similar samples in our test data we get an accuracy similar to our training accuracy, showing we have not overfit our TNN on the training and validation data. With majority voting we see that using either the majority samples or majority mean threshold method gives a very accurate score. This is perhaps expected given the training results had high accuracy and very small loss, with a high number of samples per class. This means any incorrect values would essentially be ignored as outliers. Hedges' g scores lowest for this experiment but still scores above 90\% in most cases. This shows that generally the mean similarity score is consistent across days. With a larger effect size noted, the average mean would be quite different between training and testing as noticed by Hedges' g, but not large enough to be above the 0.5 threshold.

\footnotesize
\begin{table}[H]
\centering

\begin{tabular}{l|l|l|l|l|l|l|l|l|l|l|l|}
\cline{2-12}
 &
  R1 &
  R2 &
  R3 &
  R4 &
  R5 &
  R6 &
  R7 &
  R8 &
  R9 &
  R10 &
  \multicolumn{1}{c|}{Average} \\ \hline
\multicolumn{1}{|l|}{Mean Threshold} &
  {\textbf{ 100\%}} &
   {\textbf{ 100\%}} &
  {\textbf{ 100\%}} &
   {\textbf{97.92\%}} &
  {\textbf{ 100\%}} &
  {\textbf{ 100\%}} &
   {\textbf{ 100\%}} &
   {\textbf{ 100\%}} &
  {\textbf{ 100\%}} &
   {\textbf{ 100\%}} &
   {\textbf{99.79\%}} \\ \hline
\multicolumn{1}{|l|}{Majority Samples} &
   {\textbf{ 100\%}} &
   {\textbf{ 100\%}} &
   {\textbf{ 100\%}} &
   {\textbf{97.92\%}} &
   {\textbf{ 100\%}} &
  {\textbf{ 100\%} }&
   {\textbf{ 100\%}} &
  {\textbf{ 100\%} }&
   {\textbf{ 100\%}} &
   {\textbf{ 100\%}} &
   {\textbf{99.79\%}} \\ \hline
\multicolumn{1}{|l|}{All Samples} &
  97.97\% &
  98.59\% &
  99.07\% &
  96.62\% &
  98.79\% &
  98.84\% &
  98.63\% &
  98.92\% &
  98.23\% &
  97.76\% &
  98.34\% \\ \hline
\multicolumn{1}{|l|}{Hedges} &
  89.58\% &
  93.75\% &
  95.83\% &
  91.67\% &
  87.50\% &
  \textbf{95.83\%} &
  91.67\% &
  93.75\% &
  89.58\% &
  87.50\% &
  91.67\% \\ \hline
\end{tabular}
\caption{TNN testing results across the 10 runs for our four different measures of accuracy when identifying stable versions. }
\label{tab:TNN_model_results}
\end{table}
\normalsize

\footnotesize
\begin{table}[H]
\centering
\begin{tabular}{|c|c|c|}
\hline
& TNN Worst Identification (Nanoleaf Shapes) & TNN Best Identification (Energenie) \\ \hline
Baseline Samples &   898/2
   &  900/0
   \\ \hline
   Validation Samples  &    777/123
 &  900/0
   \\ \hline
Testing Samples  &  440/460 &  887/13 \\ 
                 &  483/417 &  900/0  \\ 
                 &  862/38  &  900/0  \\ 
                 &  876/24  &  900/0  \\ \hline

Baseline Mean    & 0.0089 &  0.0001  \\ \hline
Validation Mean  & 0.1388 &  0.0002  \\ \hline
Testing Means    & 0.5187 &  0.0497  \\
                 & 0.4552 &  0.0001  \\
                 & 0.0410 &  0.0002  \\
                 & 0.0318 &  0.0001  \\ \hline

Hedges Testing Scores &  1.6676 & 0.4875 \\
                      &  1.4745 & 0.1051 \\
                      &  0.2790 & 0.5325 \\
                      &  0.2434 & -0.1263 \\ \hline

\end{tabular}
\caption{Statistics for the TNN performance across different devices.}
\label{tab:best_worst_models}
\end{table}
\normalsize

We can further analyse our results in detail. Our TNN 
performed poorly at identifying the Nanoleaf Shapes device. If we concentrate on the results from run 4, where the TNN generally performed poorly across all accuracy measures in testing (shown in Table~\ref{tab:best_worst_models}), we can note the following. In terms of the number of samples above and below the 0.5 similarity threshold, with our training, or baseline set, the model correctly identified almost all 900 samples (898), whereas in validation this number dropped to 777, and subsequently further dropped in testing where it identified 440 correctly and 460 incorrectly in the worst case. This however will only affect the `all samples' accuracy, as due to majority voting with samples, our `majority samples' accuracy would ignore those outliers. When considering the mean we see the baseline score was 0.0089, whereas in validation this rose to 0.1388, and over 0.5 in testing. This then affected the Hedges' g result, scoring 1.6676, showing a very large effect. We can also look where our TNN performed well - identifying our Energenie device. We see in both training and unseen validation data that all 900 samples are correctly scored below the 0.5 threshold. Generally this holds true for testing too, the worst performance being 887 of 900 classified correct.

We also see that the baseline, validation and testing mean are generally similar, except for that first testing day, at around 0.0001. The Hedges' g effect size is small in two cases and large in two others, which may be expected on our worst performing run. We can see that generally our technique works, although for same version identification, majority voting performs best.

\subsection{Version Change Test}
\label{sec:results_versions}

The second experiment was to use our TNN to identify when a device had changed version. Our TNN instances have only been trained on single versions per device, they have not been trained on different versions and so these are truly unknown images for the TNN to compute the similarity of. We use the same 4 accuracy measures here. Our first observation is that using a threshold of 0.5 as standard for a TNN does not work. If we use a standard threshold either with all samples, majority samples or majority mean threshold, our score is always less than the Hedges' g score, and also less than 50\% accurate. Generally the Hedges' g value is approximately 20\% more accurate than the other standard accuracy measures.

\footnotesize
\begin{table}[H]
\centering
\begin{tabular}{l|l|l|l|l|l|l|l|l|l|l|l|}
\cline{2-12}
 &
  \multicolumn{1}{c|}{R1} &
  \multicolumn{1}{c|}{R2} &
  \multicolumn{1}{c|}{R3} &
  \multicolumn{1}{c|}{R4} &
  \multicolumn{1}{c|}{R5} &
  \multicolumn{1}{c|}{R6} &
  \multicolumn{1}{c|}{R7} &
  \multicolumn{1}{c|}{R8} &
  \multicolumn{1}{c|}{R9} &
  \multicolumn{1}{c|}{R10} &
  \multicolumn{1}{c|}{Average} \\ \hline
\multicolumn{1}{|l|}{Mean Threshold} &
  34.38\% &
  25.00\% &
  25.00\% &
  46.88\% &
  25.00\% &
 34.38\% &
  50.00\% &
  37.50\% &
  46.88\% &
  28.13\% &
  35.31\% \\ \hline
\multicolumn{1}{|l|}{Majority Samples} &
  25.00\% &
  25.00\% &
  25.00\% &
  40.63\% &
  25.00\% &
  34.38\% &
  50.00\% &
  37.50\% &
  46.88\% &
  28.13\% &
  33.75\% \\ \hline
\multicolumn{1}{|l|}{All Samples} &
  32.00\% &
  27.16\% &
  29.29\% &
  44.12\% &
  31.06\% &
  35.63\% &
  48.18\% &
  31.20\% &
  45.08\% &
  37.59\% &
  36.13\% \\ \hline
\multicolumn{1}{|l|}{Hedges' g} &
   {\textbf{ \textbf{68.75\%}}} &
  {\textbf{ \textbf{65.63\%}}} &
  {\textbf{ \textbf{84.38\%}}} &
  {\textbf{ \textbf{75.00\%}}} &
  {\textbf{ \textbf{75.00\%}}} &
  {\textbf{ \textbf{62.50\%}}} &
  {\textbf{ \textbf{65.63\%}}} &
  {\textbf{ \textbf{62.50\%}}} &
  {\textbf{ \textbf{81.25\%}}} &
  {\textbf{ \textbf{71.88\%}}} &
  {\textbf{ \textbf{71.25\%}}} \\ \hline
\end{tabular}
\caption{TNN testing results across the 10 runs for our four different measures of accuracy when identifying version changes.}
\label{tab:TNN_version_results}

\end{table}
\normalsize

Further analysing the results, picking out some of the results from run 4 day 5 - our TNN performed worst at identifying the Tapo L530e device which had all samples marked as incorrect. We see that even when the version has changed, the mean for the baseline and testing is very similar, showing that any change the device has between versions is not noticeable on the wire for our technique (or possibly for any technique). Our Hedges' g score also shows a small effect size, indicating the testing images are similar. Our next observation is the LIFX A19 which only had a subtle change to the TNN. We see here that using the samples accuracy metric, only 72 of the 900 are scored as above 0.5, and therefore classed as different, which is a poor score. This is further shown by the testing mean being 0.833, which, although higher than the baseline of 0.00003, is not high enough to be classed by a standard TNN as different enough. However, when we use Hedges' g, we see the change in mean is classed as significant - an effect size of over 0.5. For this set of data, Hedges' g is therefore the best measure to use. This is likely to be the reason for the 20\% difference in accuracy scores in Table~\ref{tab:TNN_version_results}. Our final observation is for devices that have changed by a large amount. Our TNN scored highly on all accuracy metrics when identifying the TPLink HS110 device. The mean had gone from 0.0006 to 0.9960, showing a high difference in similarity. This aligns to both the samples, which all were classed as 'dissimilar' and the Hedges' g score being over 66 (when our threshold for difference is only 0.5). We can conclude the following: Hedges' g can be the most useful accuracy metric for version changes, as it can pick up both large differences and subtle differences, and is shown to be, on average, 20\% more accurate than the standard TNN measure. Hedges' g is therefore needed for version change identification as some device version changes are hugely different when others are subtle. This is therefore potentially a solution to the problem of identifying device version changes.

\small
\begin{table}[H]
\centering
\begin{tabular}{|c|c|c|c|}
\hline
& Worst Performing (Tapo L530e) & Subtle change (LIFX A19) & Best performing (TP-Link HS110) \\ \hline
Baseline Samples &  900/0  &  900/0   &  900/0   \\ \hline
Testing Samples  &  900/0  &  828/72  &  0/900   \\ \hline
Baseline Mean    &  0.0005 &  0.00003 &  0.0006  \\ \hline
Testing Mean     &  0.0010 &  0.0833  &  0.9960  \\ \hline
Hedges Score     &  0.0761 &  0.5349  &  66.7936 \\ \hline
\end{tabular}
\caption{Statistics for the worst, subtle change and best performing examples regarding version changes.}
\label{tab:best_worst_versions}
\end{table}
\normalsize

\section{Results Analysis and Discussion}
\label{sec:discussion}

We will now discuss the results from the previous section in more detail and analyse possible reasons for the results, again discussing generally across the 10 runs.

\subsection{Training and Validating the TNNs}

\begin{figure}[H]
    \centering
    \includegraphics[width=1\linewidth]{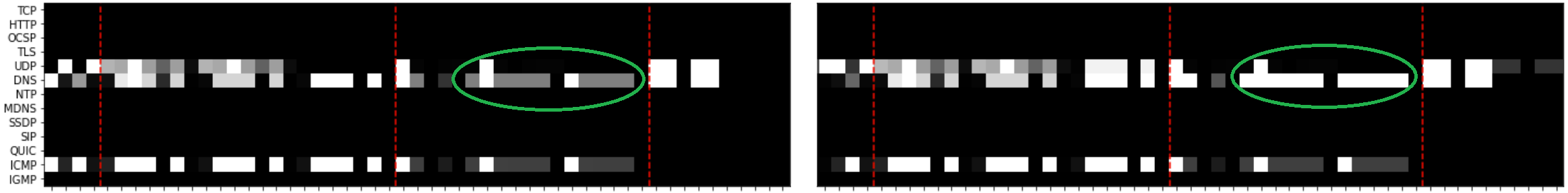}
    \caption{Visualisation of Energenie similar training pairs.}
    \label{fig:nn-similar-pairs-energenie}
    \Description[2 images of similar training pairs.]{Visualisation of Energenie similar training pairs.}
\end{figure}

When training our TNN, it was shown than our input data was relatively balanced - all but 1 device had the full 900 samples available for testing, and all but 2 had the full 900 samples available for validation. We also saw that training and validation yielded similar results of a high accuracy of over 99\% in cases of all samples (dissimilar and similar) and similar only. When looking at the similar input images for training, we can see why our technique of using a TNN with image similarity works and produces a high accuracy score. Figure~\ref{fig:nn-similar-pairs-energenie} shows one of the training pairs for the Energenie device. The figure shows the images are similar in the sense that only 3 protocols are shown to be in use - UDP, DNS and ICMP. The green circles show the subtle differences between the images - within the `time period' features in the left hand side image have more variance between the values, noted by the change in greyscale, whereas the right hand side image values are all approximately the same as they are all white. These subtle differences are what the TNN uses to identify devices, and why our mean similarity score was not exactly 0 (and is very unlikely to be).

\begin{figure}[H]
    \centering
    \includegraphics[width=1\linewidth]{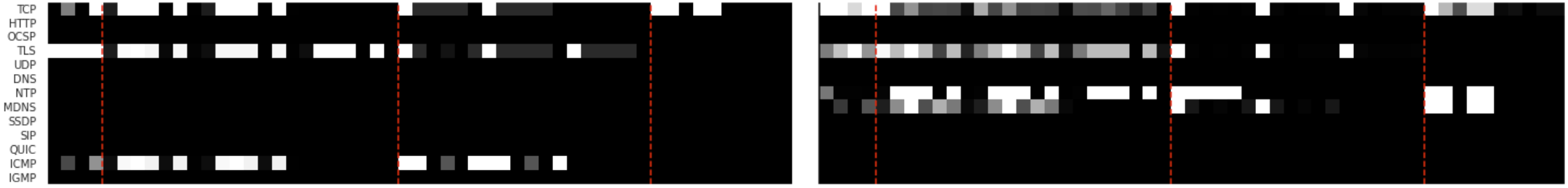}
    \caption{visualisation of Tapo L530e vs Nanoleaf Shapes dissimilar training pairs.}
    \Description[tapo vs Nanoleaf dissimilar training pairs]{visualisation of Tapo L530e vs Nanoleaf Shapes dissimilar training pairs.}
    \label{fig:nn-dissimilar-pairs-tapo}
\end{figure}

In terms of dissimilar images, we needed to ensure that the images were clearly dissimilar between devices. In Fig.~\ref{fig:nn-dissimilar-pairs-tapo} the Tapo L530e and Nanoleaf Shapes devices do produce quite dissimilar images with features that the TNN would identify. Both devices used TCP and TLS protocols which had varying packet rates and magnitudes, however the Tapo L530e device (LHS of Fig.~\ref{fig:nn-dissimilar-pairs-tapo}) had ICMP packets whereas the Nanoleaf Shapes device (RHS of Fig.~\ref{fig:nn-dissimilar-pairs-tapo}) uses NTP and mDNS more frequently. If we compare these two images to the ones for the Energenie device in Fig.~\ref{fig:nn-similar-pairs-energenie} we can see the images are all distinct per device.

\begin{figure}[H]
    \centering
    \includegraphics[width=1\linewidth]{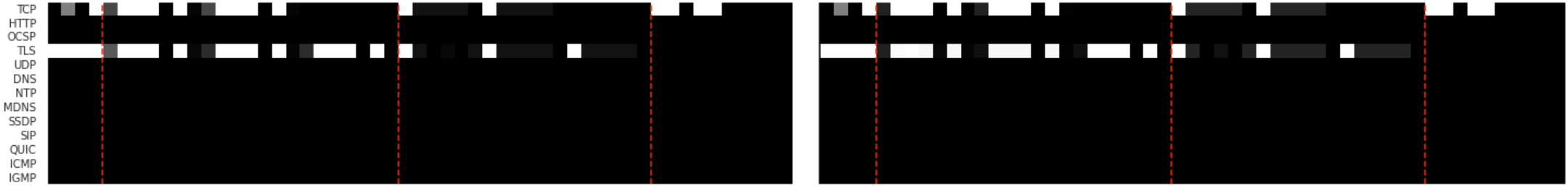}
    \caption{An example false positive - the two different device images (Dyson Fan and Tapo L530e) were erroneously classed by the TNN as similar.}
    \Description[]{}
    \label{fig:nn-training-fp}
\end{figure}

\begin{figure}[H]
    \centering
    \includegraphics[width=1\linewidth]{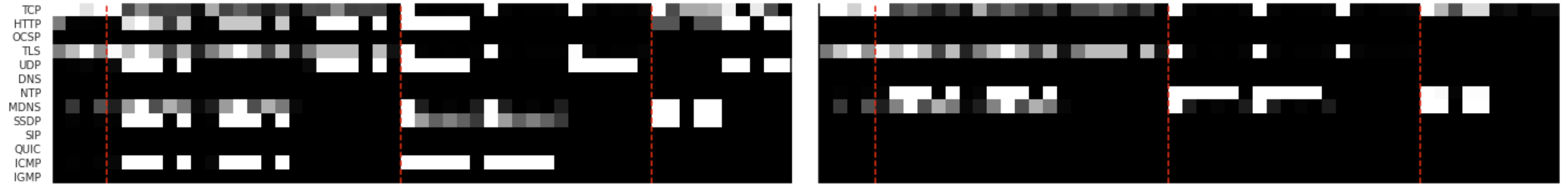}
    \caption{An example false negative - the two different device images from the same device (Nanoleaf Shapes) were erroneously classed by the TNN as dissimilar.}
    \Description[False negative.]{An example false negative - the two different device images from the same device (Nanoleaf Shapes) were erroneously classed by the TNN as dissimilar.}
    \label{fig:nn-training-fn}
\end{figure}

We did notice a small number of mis-classifications across training and validation, both false positive, where dissimilar samples were classed as similar, and false negative, where similar samples were classed as dissimilar. Starting with the false positive shown in Fig.~\ref{fig:nn-training-fn}, we see for this sample that the Dyson Fan device (LHS of Fig.~\ref{fig:nn-training-fp}) and the Tapo L530e device (RHS of Fig.~\ref{fig:nn-training-fp}) do indeed look very similar, with our technique outputting a score of 0.021 for these images, which aligns to the fact that some pixels have a slightly different colour. We can see however that these cases didn't affect the technique much, due to this only being a very small number of samples in both training and validation. For the false negative, we have samples from the Nanoleaf Shapes device, which the TNN does not perform well at identifying in either training or validation. We can see in Fig.~\ref{fig:nn-training-fn} how this device does not have an overly consistent fingerprint. For the TCP, TLS and mDNS protocols they look vaguely similar across samples. However the first sample also uses more uDP, SSDP and ICMP, making it look quite different, noted by the score given of 0.579, just over the threshold of being similar.

We have shown that our TNN produces a high accuracy score during training and validation and from analysis of the input images we can see why this is the case. Across our two days for training the images seem to be similar enough between sample periods for a device and dissimilar enough between different devices to distinguish them. We also explored the main misclassifications but note that they didn't affect the overall TNN accuracy by much. One thing worth exploring further here is around tweaking the TNN for potentially better results, for example if day 3 fails to get a high enough validation accuracy. We will expand on this further in Section~\ref{sec:realworld}.

\subsection{Stable Version Test}

\begin{figure}[H]
    \centering
    \includegraphics[width=1\linewidth]{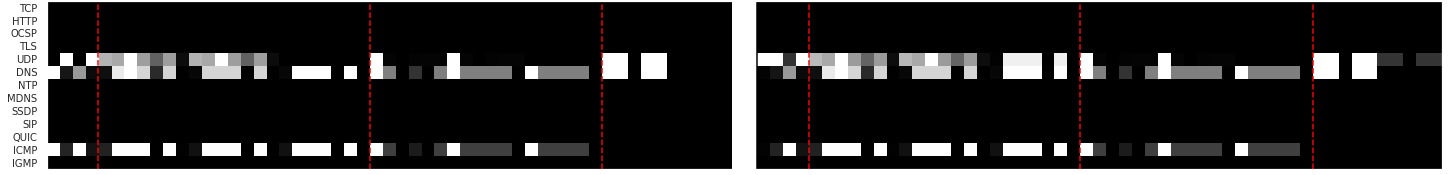}
    \caption{Visualisation of Energenie device similar training pairs.}
    \Description[Similar training pairs.]{Visualisation of Energenie device similar training pairs.}
    \label{fig:nn-testing-model-energenie}
\end{figure}

\begin{figure}[H]
    \centering
    \includegraphics[width=1\linewidth]{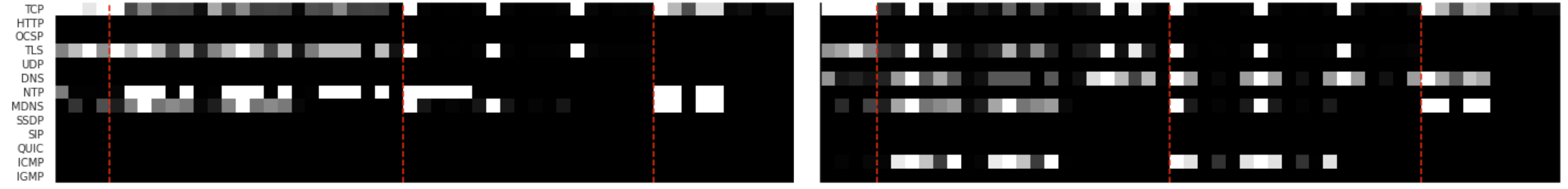}
    \caption{Visualisation of Nanoleaf Shapes device similar training pairs.}
    \Description[Nanoleaf similar training pairs.]{Visualisation of Nanoleaf Shapes device similar training pairs.}
    \label{fig:nn-testing-model-nanoleaf}
\end{figure}

We had four days worth of data to test our TNN at identifying devices with a stable version. In the results section we saw that our technique performed well for the Energenie device. If we look at the input images in Fig.~\ref{fig:nn-testing-model-energenie} we can see why this is the case. The images shown are almost identical in terms of protocols used and magnitudes, so it makes sense than the TNN worked well in this case. We did however, again as expected, see that the TNN struggled with identification of the Nanoleaf Shapes device. If we look at those input images, specifically one that was misclassified in Fig.~\ref{fig:nn-testing-model-nanoleaf} we can see that the issue is not the trained TNN as it work as expected in terms of identifying that the images are different, the issue is that that images coming from the same device are quite different. We highlight some of the main differences between these two images, namely more protocols being used and magnitude of pixels being vastly different. We have shown however that our technique does work for device stable version identification, with any of the accuracy scores for the samples being useful as a metric.


\subsection{Version Change Test}

\begin{figure}[H]
    \centering
    \includegraphics[width=1\linewidth]{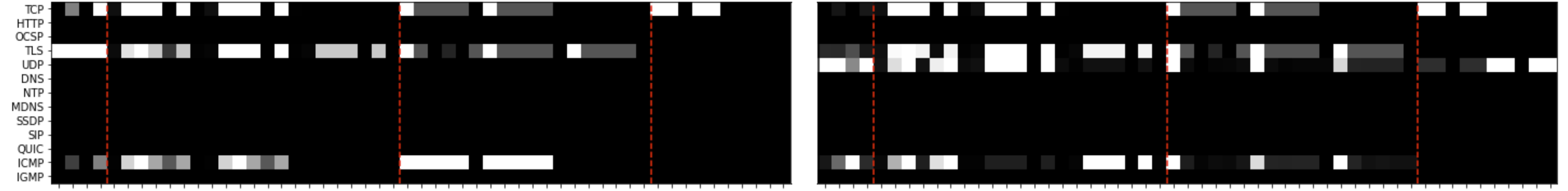}
    \caption{TPLink HS110 input images for v1.0.10 (LHS) and v1.2.6 (RHS).}
    \Description[HS110 version change images.]{TPLink HS110 input images for v1.0.10 (LHS) and v1.2.6 (RHS).}
    \label{fig:hs110_version_images}
\end{figure}

\begin{figure}[H]
    \centering
    \includegraphics[width=1\linewidth]{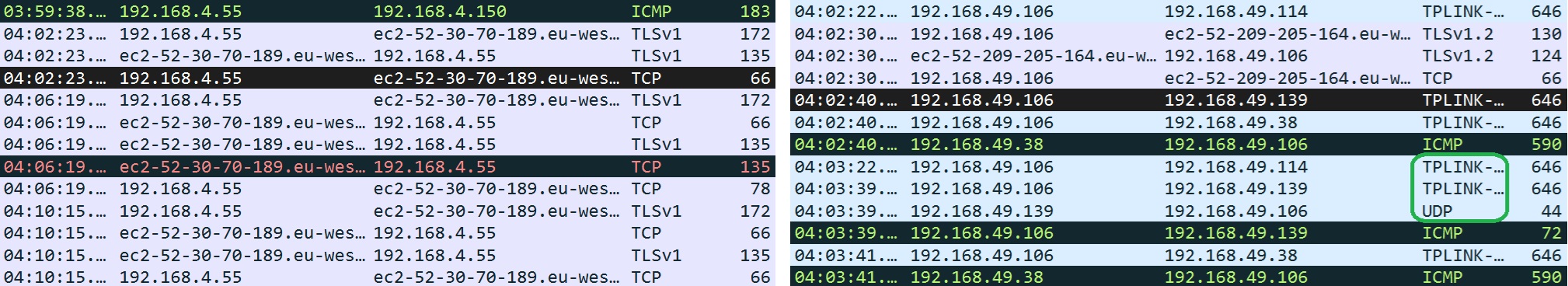}
    \caption{Wireshark capture of the TPLink HS110 device v1.0.10 (LHS) and v1.2.6 (RHS).}
    \Description[HS110 Wireshark.]{Wireshark capture of the TPLink HS110 device v1.0.10 (LHS) and v1.2.6 (RHS).}
    \label{fig:hs110_version_wireshark_preupdate}
\end{figure}

We now discuss the version change results, exploring further the results shown in Section 5.3. As mentioned, these images were unknown to the TNN during the training phase. Starting with a device that showed a large, obvious change (TPLink HS110), we see in Fig.~\ref{fig:hs110_version_images} that the pattern is similar for TCP, TLS and ICMP traffic, however after the update more UDP traffic was seen. If we then compare the PCAPs of pre and post update (Fig.~\ref{fig:hs110_version_wireshark_preupdate}) we can see the main changes align. There is still TCP and TLS traffic however the number of bytes post update for these protocols has decreased. We also note that a new `TPLINK' protocol is in use, which is based on UDP.

\begin{figure}[H]
    \centering
    \includegraphics[width=1\linewidth]{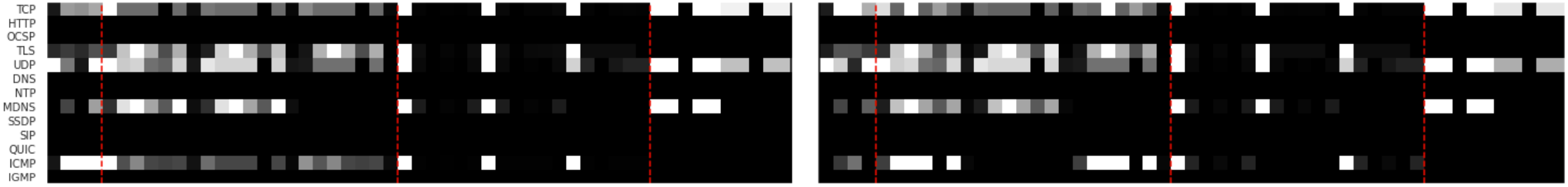}
    \caption{LIFX A19 input image comparison between v2.7 (LHS) and v2.8 (RHS).}
    \Description[A19 input images.]{LIFX A19 input image comparison between v2.7 (LHS) and v2.8 (RHS).}
    \label{fig:lifx_version_images}
\end{figure}

\begin{figure}[H]
    \centering
    \includegraphics[width=1\linewidth]{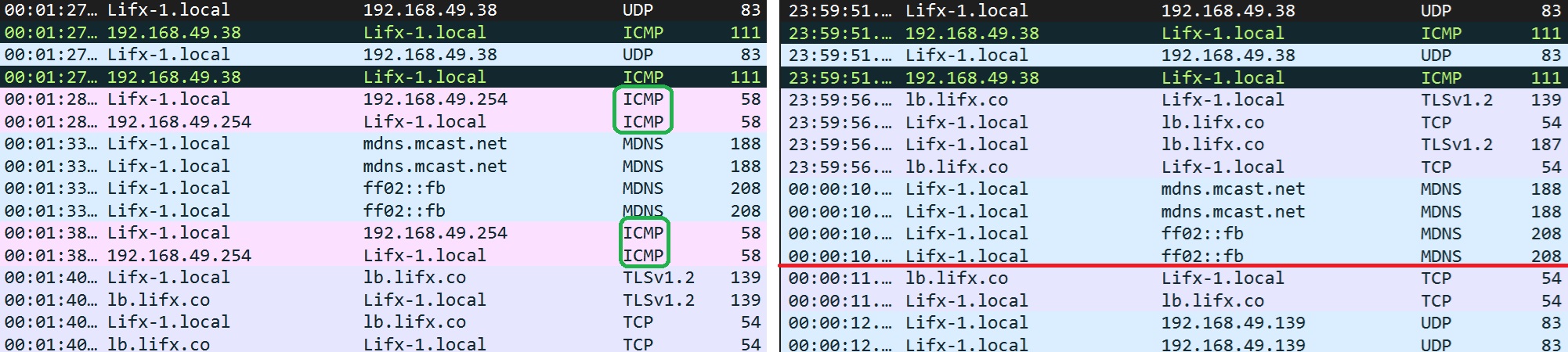}
    \caption{Comparison of LIFX A19 v2.7 (LHS) and v2.8 (RHS) firmware version traffic captures.}
    \Description[LIFX Wireshark images.]{Comparison of LIFX A19 v2.7 (LHS) and v2.8 (RHS) firmware version traffic captures.}
    \label{fig:lifx_version_preupdate}
\end{figure}

We now look at the case where there were subtle differences picked up between device firmware versions, where Hedges' g was needed over standard thresholds to notice this change. For this we saw the LIFX A19 device had subtle differences available between versions, picked up by the TNNs shown in Fig.~\ref{fig:lifx_version_images}. Fig.~\ref{fig:lifx_version_preupdate} shows the capture before and after the update. If we analyse the Wireshark captures between versions at approximately midnight when our first sample would be taken, we see that the LIFX device does broadly the same things for 12 packets - it has UDP, mDNS, TLS and TCP packets of the same size, plus a device on the local network pings the LIFX device. However we notice that post-update the LIFX device no longer pings the gateway (highlighted green in the figure) and just gets periodic pings to it. Our choice of devices meant that devices on the network would ping each other, but this is ok for two reasons. Even without them the LIFX version change would still be noticed as the LIFX stopped pinging the gateway, plus in a deployment the learned behaviours of the network would contribute to the baseline normal. This change aligns to the TNN greyscale image - prior to the update it was more balanced with grey pixels for server and client, after the update it became white and related to the server only.

\begin{figure}[H]
    \centering
    \includegraphics[width=1\linewidth]{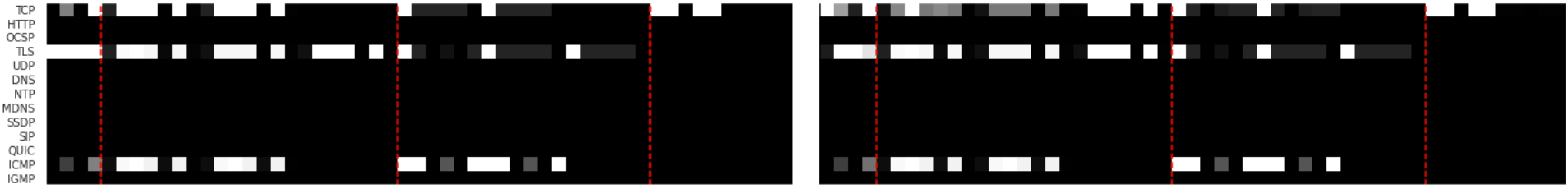}
    \caption{Tapo L530e device showing no difference in input images across versions 1.2.2 (LHS) and 1.2.4 (RHS).}
    \Description[Tapo with no version changes.]{Tapo L530e device showing no difference in input images across versions 1.2.2 (LHS) and 1.2.4 (RHS).}
    \label{fig:tapo_version_nochange}
\end{figure}

\begin{figure}[H]
    \centering
    \includegraphics[width=1\linewidth]{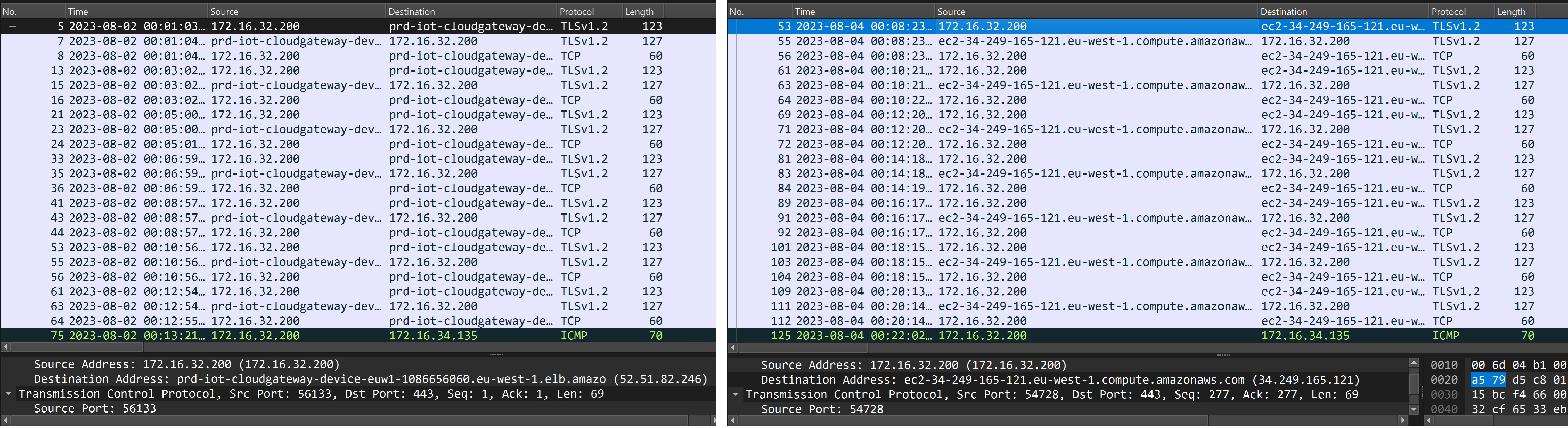}
    \caption{Tapo L530e device showing no on-wire differences between version 1.2.2 (LHS) and 1.2.4 (RHS).}
    \Description[Tapo wireshark without changes.]{}
    \label{fig:tapo_wireshark_2ndaug}
\end{figure}

For all 10 trained TNNs, The Tapo L530e version change did not get identified. Starting with the input images, taking Fig.~\ref{fig:tapo_version_nochange} as an example, we see that they are essentially identical; there is no obvious difference between the two versions, which aligns to the average mean and Hedges' g score in Section~\ref{sec:results_versions}. If we then look at part of the Wireshark captures that relate to these images (Fig.~\ref{fig:tapo_wireshark_2ndaug}), we can see why this is the case. In both cases there is an exchange of 21 packets across TCP and TLS, followed by a (local) ICMP packet. These packets are all shown to be the same size, of 123, 127 and 60 bytes for the TCP and TLS exchange and 70 bytes for the ICMP request. The only differences between the versions is the source port number which appears random as expected, and the DNS hostname, but this is not a feature considered with flow techniques, and so no technique using flow statistics would work here. A keywords approach~\cite{AndrewsFMEC,AndrewsCPSIOTSEC} however might pick up this change if it was significant.

It is worth stating that on-wire changes may not be directly related to change in the firmware between versions. During a device update a device may reboot and refresh any internal configuration, or re-advertise itself on the network, making devices interact again. This is still useful however, as identifying this can also lead towards anomaly detection techniques for traffic between devices.



    
    

    




\subsection{Proposed Real World Architecture}

\label{sec:realworld}

\begin{figure}[H]
    \centering
    \includegraphics[width=0.98\linewidth]{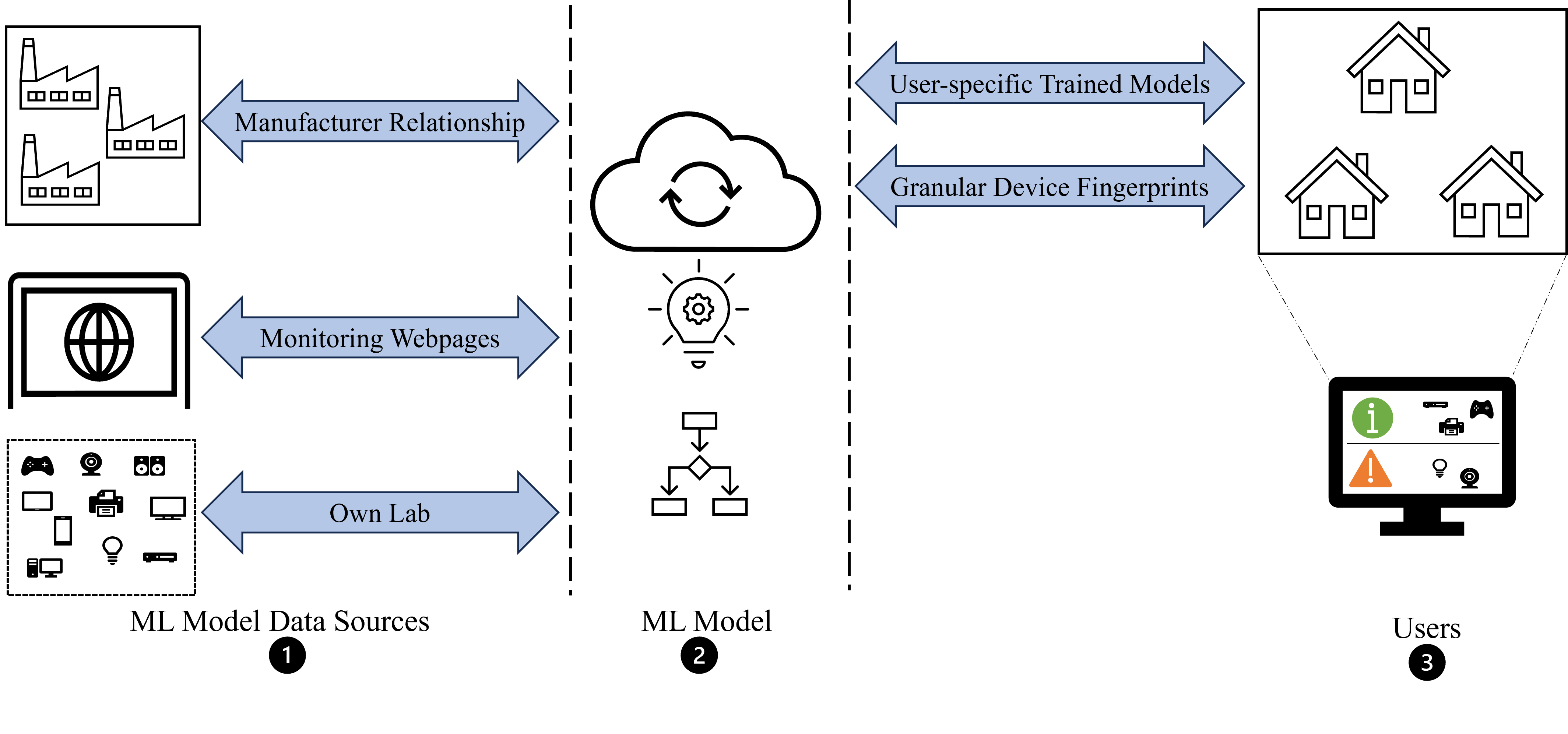}
    \caption{Graphical depiction of a real-world tool for automated version identification.}
    \Description[Real-world tool diagram.]{Graphical depiction of a real-world tool for automated version identification.}
    \label{fig:real-world}
\end{figure}

We have shown that we have a viable technique that can identify when a device is on a stable version, and also when it changes version. This technique can also be expanded to identify device models and perform anomaly detection. It is worth discussing how our technique could be enhanced to be deployed for real-world use. Our technique could be deployed as a system in the following way as shown in Fig.~\ref{fig:real-world}. We propose a cloud-based approach where a solution in the cloud would have access to all device fingerprints, and a constantly updating database of latest versions of devices. This could either be from relationships with manufacturers, monitoring websites or obtaining devices and manually checking versions. This then allows a large database of fingerprint images of what the devices should look like under normal use, running a specific firmware version, as a baseline to compare images from other networks against. This cloud-based approach would then allow fingerprints per device to be pushed down to users, to train their own ML model specific to their network. This allows the model to use both the known good fingerprints from the cloud plus any nuances to their home network, for example how devices on the LAN interact with each other. This trained model could then form the basis of an app, with results specific to each user. The tool would work in the following way:

\begin{itemize}
    \item If no change in device behaviour is detected, the device is presumed to be up to date.
    \item If change is detected, the cloud service is queried.
    \item If the response of this query says that there is a new latest version available for the device, it is presumed that the device must have updated. The model is therefore re-trained with new traffic. This can optionally be flagged to the user to confirm.
    \item If there is no new version available, the device behaviour is considered anomalous, so manual checking of the device and its traffic is required. This would be flagged to the user.
    \item If this manual checking finds a genuine compromise, the device is disabled and fixed/permanently removed; otherwise the model is re-trained.
\end{itemize}

This cloud based approach is similar to the products (Fingbox, Bitdefender BOX2) mentioned in~\cite{AndrewsWFIOT}, but our technique could enhance their capabilities. To align with other products available, such as Amazon IoT Device Defender~\cite{AmazonWebServices}, or Microsoft Defender for IoT~\cite{Microsoft} we would want to implement model re-training. Our current method uses days 1 and 2 data for training and day 3 for validation. If it was found that day 3 did not validate the model well (anomalies were detected erroneously) we could consider using more data to improve the model. For example we could use day 4 for validation against training on days 1 and 2, or re-training on days 2 and 3 with validation on day 4. We could also implement something like a `7 day rolling score' with a report to users on which days had been flagged as anomalous. It may be the case that single days were anomalous and then normal again, and it would be up to the user to look into this further if they wish. This still provides an efficiency gain for users, notifying them of a small number of devices that need checking.

\section{Conclusions and Future Work}
\label{sec:conclusions}

In this paper we identified two challenges for identifying IoT firmware versions: the on-wire behaviour resulting from different firmware versions running on the same device are more subtle than those for different device models, types or manufacturers; and there is limited data on different device versions due to lack of publicly available datasets. These challenges are novel in this area and have therefore not previously been addressed by the literature.
We developed a technique using flow statistics to transform on-wire device behaviours into greyscale images. These images were fed into a Twin Neural Network model to output similarity scores. Our best performing model was 95.83\% accurate at identifying stable versions and 84.38\% accurate identifying version changes. 
By calculating the Hedges' g effect size of similarity scores, we were able to detect the subtle changes resulting from a device running different firmware versions. We showed how Hedge's g as a metric is approximately 20\% more accurate than the standard TNN measure. By using a TNN model trained on device differences --- but tested on version changes --- we were able to overcome the limited data problem via transfer learning. This allowed the TNN to be able to correctly calculate the similarity of images that were truly unknown - the images of different device versions had not been used to train the TNN.

To improve our technique, further work could be carried out in this area, including:



\begin{enumerate}
\item Using active approaches. Although stated to not be ideal for the reasons mentioned in the introduction, active approaches may have a place in augmenting techniques when identifying device versions to overcome the problem when a device has no change in passive on-wire signature. This could either be from banner grabbing on known ports (as the version change may update versions or banners of extra components such as libraries), or logging into device management pages and identifying version strings.
\item Using this technique to identify newly added devices, or expanding to a huge database of known devices as an identification technique for unknown devices.
\item We have shown our technique can identify anomalies in terms of version changes, there would be value in applying the same technique for other types of anomaly detection from compromised devices.
\end{enumerate}

\bibliographystyle{ACM-Reference-Format}
\bibliography{sample-base}

\appendix
\section{Extended Stable Version Results}
  \begin{table}[H]
    \centering
    \caption{Number of samples that are below (left of bracket) and above (right of bracket) the 0.5 threshold for each of the 10 runs over each of the four days per device.}
    \includegraphics[width=0.75\linewidth]{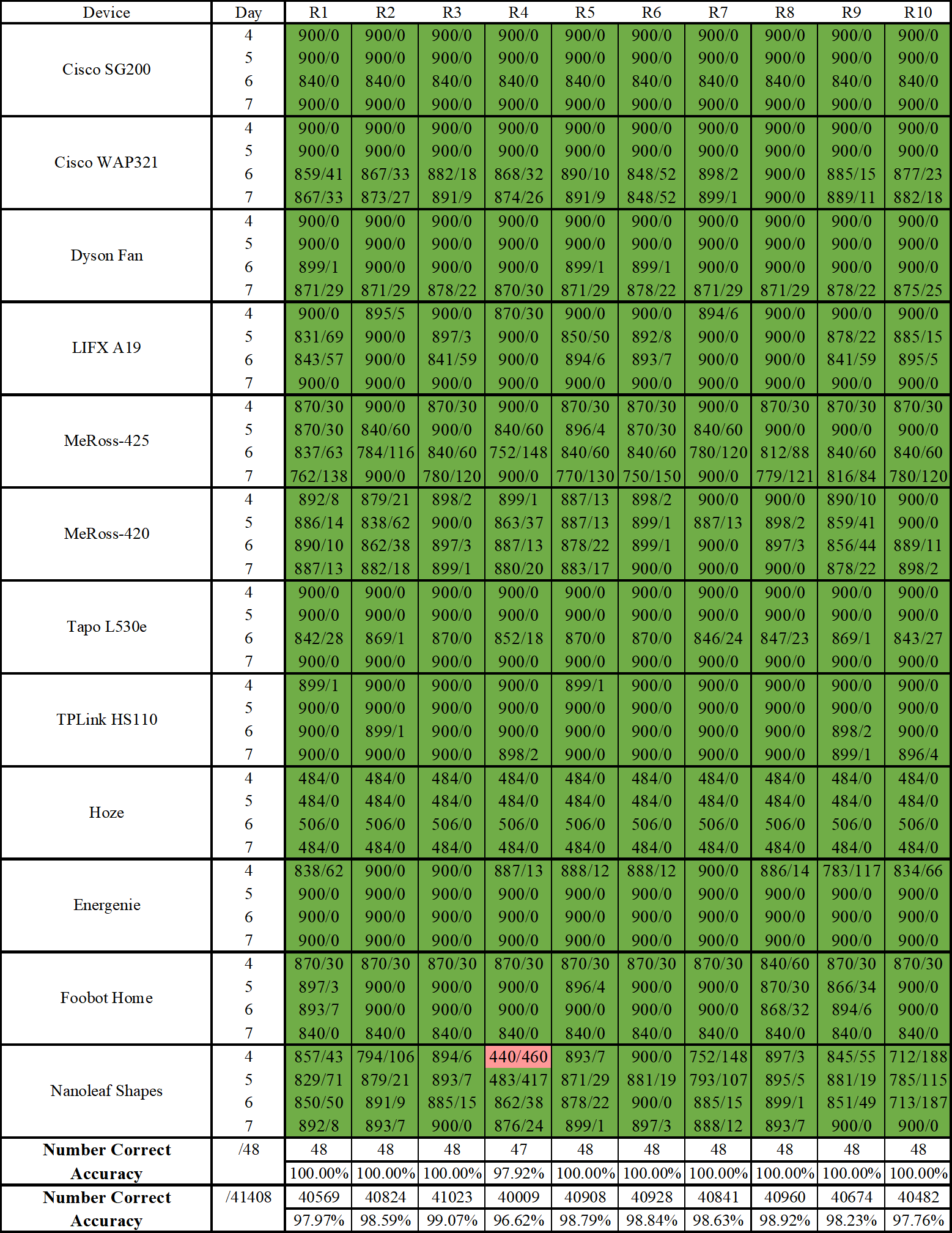}
    \label{fig:app-mod-samples}
 \end{table}
 
 \begin{table}[H]
    \centering
    \caption{Mean similarity scores for each of the 10 runs over each of the four days per device.}
    \includegraphics[width=0.75\linewidth]{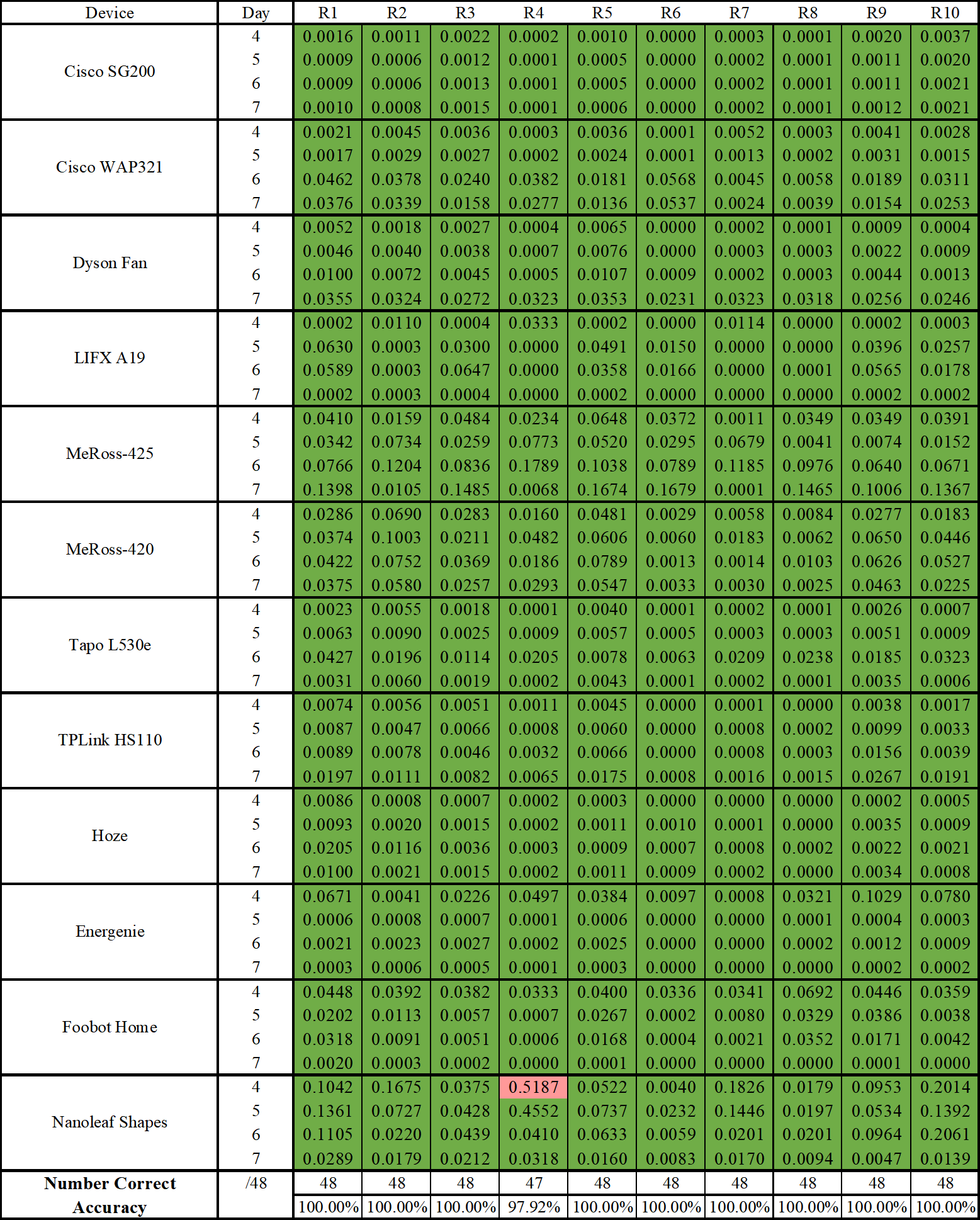}
    \label{fig:app-mod-mean}
 \end{table}

  \begin{table}[H]
    \centering
    \caption{Hedges' g scores for each of the 10 runs over each of the four days per device.}
    \includegraphics[width=0.75\linewidth]{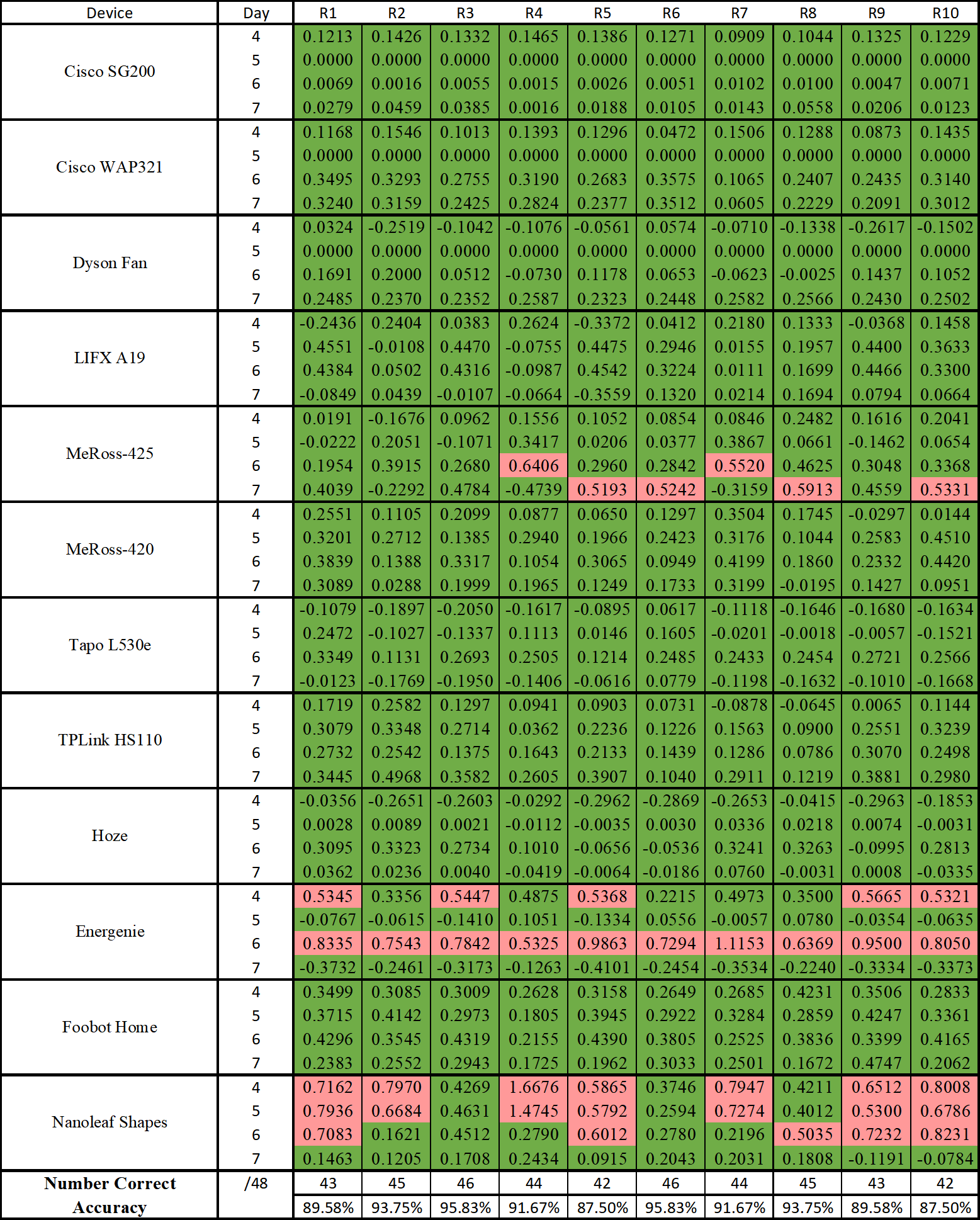}
    \label{fig:app-mod-hedges}
 \end{table}

\section{Extended Version Change Results}

\begin{table}[H]
    \centering
    \caption{Mean similarity scores for each of the 10 runs over each of the four days per device.}
    \includegraphics[width=0.73\linewidth]{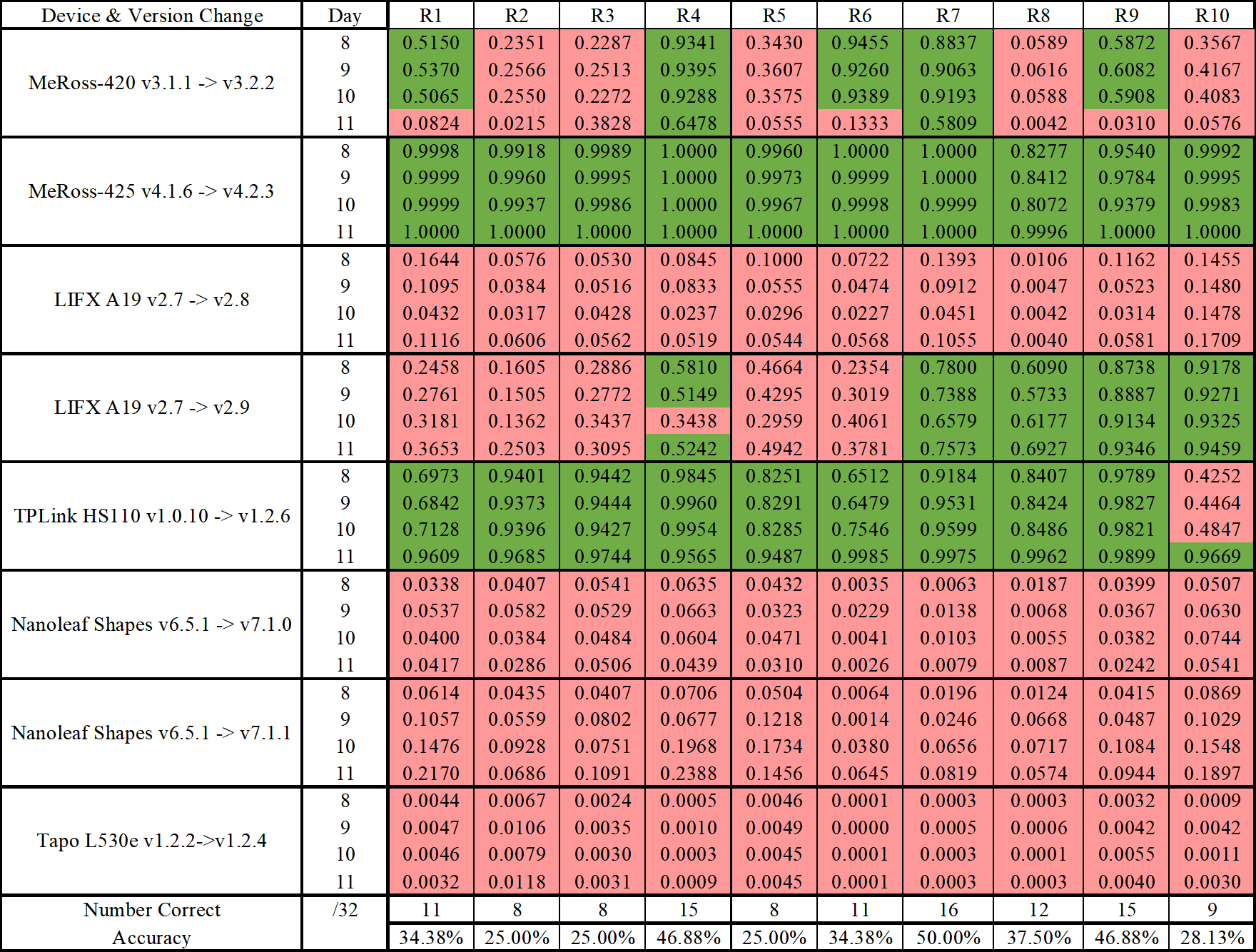}
    
    \label{fig:app-vers-mean}
\end{table}

\begin{table}[H]
    \centering
    \caption{Hedges' g scores for each of the 10 runs over each of the four days per device.}
    \includegraphics[width=0.73\linewidth]{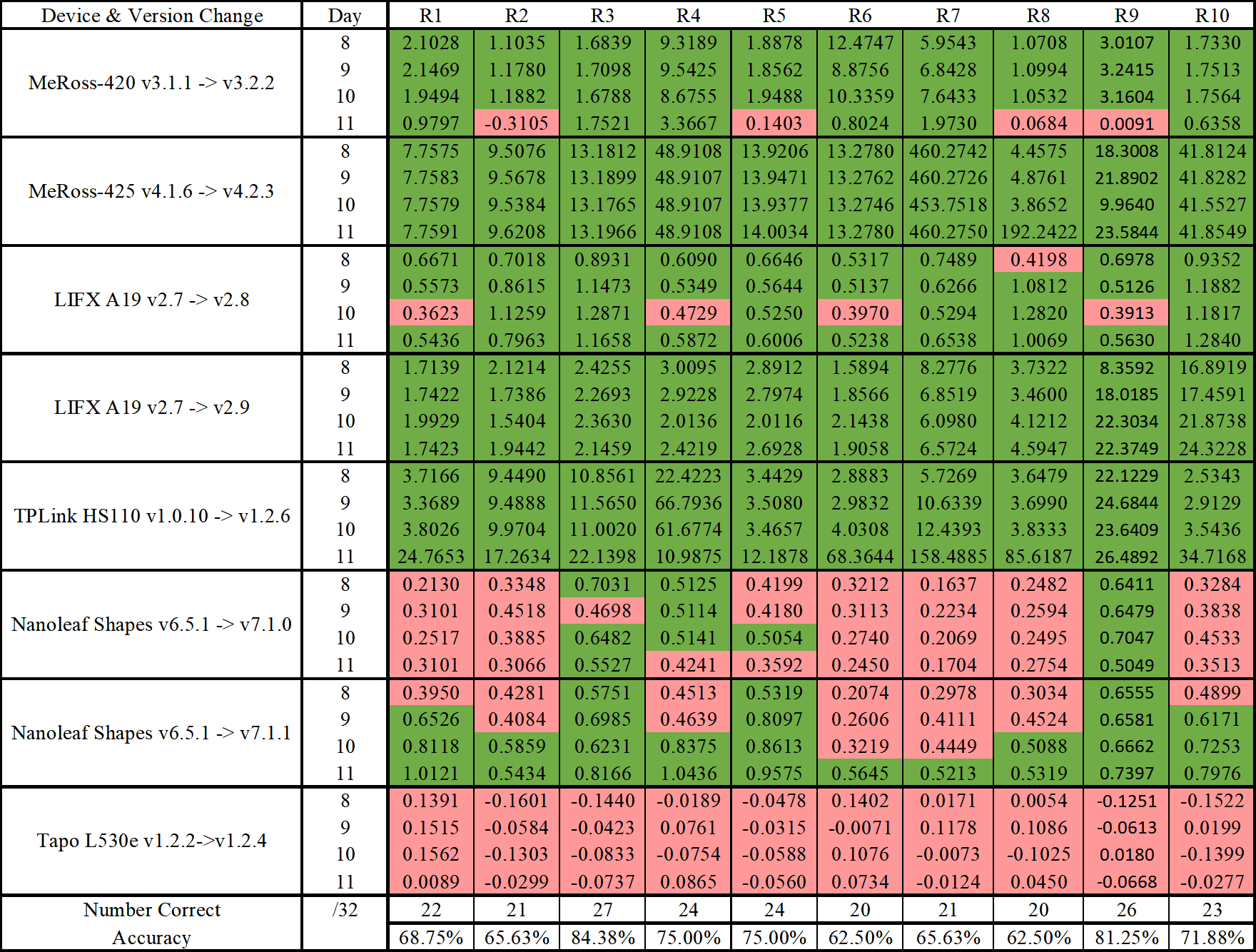}
    
    \label{fig:app-vers-hedges}
\end{table}

\begin{table}[H]
    \centering
    \caption{Number of samples that are below (left of bracket) and above (right of bracket) the 0.5 threshold for each of the 10 runs over each of the four days per device.}
    \includegraphics[width=0.73\linewidth]{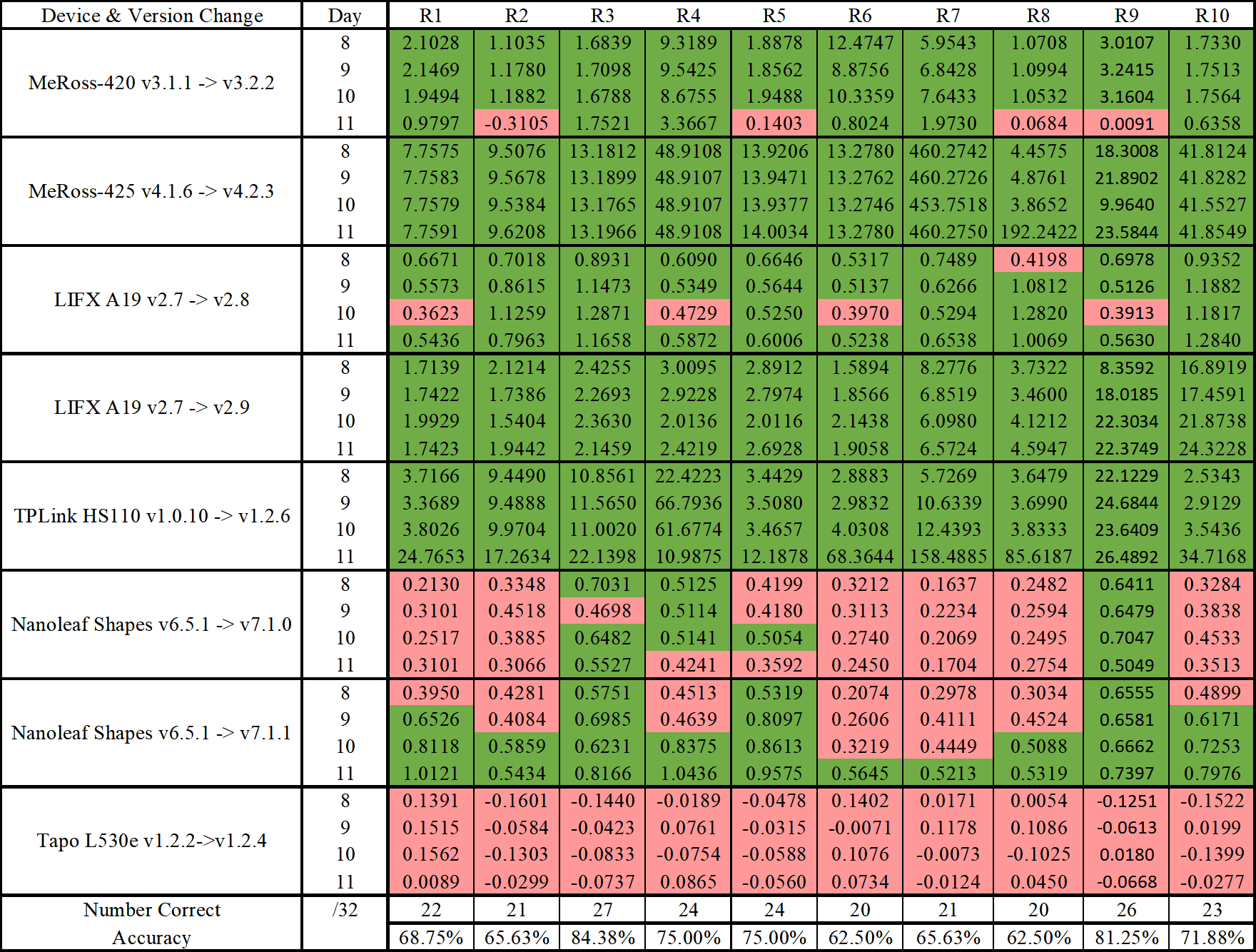}
    
    \label{fig:app-vers-samples}
\end{table}

\end{document}